\documentclass[a4paper,11pt]{article}
\pdfoutput=1 % if your are submitting a pdflatex (i.e. if you have
\usepackage{jheppub} % for details on the use of the package, please
\usepackage{enumitem}

\usepackage[T1]{fontenc} % if needed

\usepackage{braket}
\usepackage[dvipsnames]{xcolor}
\usepackage{comment}
\usepackage{float}
\usepackage{caption}
\usepackage{subcaption}
\usepackage{bbold}  
\usepackage{hyperref}

\newcommand{\al}{{\alpha}}
\newcommand{\bt}{{\beta}}
\newcommand{\de}{{\delta}}

\newcommand{\lm}{{\lambda}}

\newcommand{\sig}{{\sigma}}
\newcommand{\vep}{{\varepsilon}}
\newcommand{\vphi}{{\varphi}}

\newcommand{\Ga}{{\Gamma}}

\newcommand{\Om}{{\Omega}}

\newcommand{\bfR}{{\mathbf{R}}}

\newcommand{\rB}{{\mathrm{B}}}
\newcommand{\rE}{{\mathrm{E}}}

\newcommand{\pd}{{\partial}}

\usepackage{amsmath}

\title{\boldmath Thin-wall vacuum decay in the presence of a  compact dimension meets the $H_0$ and $S_8$ tensions}

\author[a,b,c]{Luis A. Anchordoqui}
\author[d,e]{Ignatios Antoniadis}
\author[d]{Daniele Bielli}
\author[d]{Auttakit Chatrabhuti}
\author[d]{Hiroshi Isono}

\affiliation[a]{Department of Physics and Astronomy, Lehman College, City University of New York, NY 10468, USA}
\affiliation[b]{Department of Physics, Graduate Center, City University of New York, NY 10016, USA}
\affiliation[c]{Department of Astrophysics, American Museum of Natural History, NY 10024, USA}
\affiliation[d]{High Energy Physics Research Unit, Faculty of Science,
Chulalongkorn University,
\\
Bangkok 10330, Thailand}
\affiliation[e]{Laboratoire de Physique Th\'eorique et Hautes Energies - LPTHE
\\
Sorbonne Universit\'e, CNRS, 4 Place Jussieu, 75005 Paris, France}

\emailAdd{luis.anchordoqui@gmail.com, antoniad@lpthe.jussieu.fr, d.bielli4@gmail.com, auttakit.c@chula.ac.th, hiroshi.isono81@gmail.com}

\abstract{The proposal of a rapid sign-switching cosmological constant in the late universe, mirroring a transition from anti-de Sitter (AdS) to de Sitter (dS) space, has significantly improved the fit to observational data and provides a compelling framework for ameliorating major cosmological tensions, such as the $H_0$ and $S_8$ tensions. An attractive theoretical realisation that accommodates the AdS $\to$ dS transition relies on the Casimir forces of fields inhabiting the bulk of a 5-dimensional (5-dim) set up. Among the fields characterising the dark sector, there is a real scalar field $\phi$ endowed with a potential holding two local minima with very small difference in vacuum energy and bigger curvature (mass) of the lower one. Shortly after the false vacuum tunnels to its true vacuum state, $\phi$ becomes more massive and its contribution to the Casimir energy becomes exponentially suppressed. The tunneling process then changes the difference between the total number of fermionic and bosonic degrees of freedom contributing to the quantum corrections of the vacuum energy, yielding the AdS $\to$ dS transition. We investigate the properties of this theoretical realisation to validate its main hypothesis and characterise free parameters of the model. We adopt the Coleman-de Luccia formalism for calculating the transition probability within the thin-wall approximation. We show that the Euclidean bounce configuration that drives the transition between $\phi$ vacua has associated at least a sixth order potential. We also show that distinctive features of the required vacuum decay to accommodate the AdS $\to$ dS transition are inconsistent with a 5-dim non-compact description of the instanton, for which the bounce is $O(5)$ symmetric, and instead call for a 5-dim instanton with a compact dimension, for which the bounce is $O(4)\times U(1)$ symmetric.}

\begin{document} 
\maketitle
\flushbottom

\section{Introduction}
\label{sec:1}

Over time and through many experiments, $\Lambda$-cold-dark-matter ($\Lambda$CDM) has become established as a well-tested phenomenologically concordant model of modern cosmology, accommodating simultaneously
data from the cosmic microwave background (CMB), baryon acoustic oscillation (BAO), and type Ia supernovae (SNe Ia)~\cite{ParticleDataGroup:2024}. However,  precision data from recent experiments were able to pierce this concordant model's resistant armor~\cite{Abdalla:2022yfr}. In particular, $\Lambda$CDM has been cracked by an ever-enlarging ($\sim 5\sigma$) tension on the Hubble constant $H_0$ between the global fitting of CMB observations with $\Lambda$CDM extrapolation~\cite{Planck:2018vyg} and the local quasi-direct measurements from SNe Pantheon+  compilation~\cite{Scolnic:2021amr} with Supernovae and $H_0$ for the Equation of State of dark energy (SH0ES) calibration~\cite{Riess:2021jrx,Murakami:2023xuy}. Low- and high-redshift observations have also set off a tension in the determination of the amplitude of the matter clustering in the late Universe (parametrised by $S_8$). More categorically, the value of $S_8$ inferred from Planck's CMB data assuming $\Lambda$CDM~\cite{Planck:2018vyg} is in $\sim 3\sigma$ tension with the determination of $S_8$ from the cosmic shear data of the Kilo-Degree Survey (KiDS-1000)~\cite{Heymans:2020gsg}. There has been an explosion of creativity to resolve the cosmic conundrum~\cite{DiValentino:2021izs, Schoneberg:2021qvd,Perivolaropoulos:2021jda}; the good news is that all possible explanations involve new physics. There is no feeling though that we are dotting the i's and crossing the t's of any mature theory.

$\Lambda_s$CDM~\cite{Akarsu:2019hmw, Akarsu:2021fol,Akarsu:2022typ,Akarsu:2023mfb} is one of the many new physics models that have been proposed to simultaneously resolve the $H_0$ and $S_8$ tensions. The model relies on an empirical conjecture, which postulates that $\Lambda$ may have switched sign (from negative to positive) at critical redshift $z_c \sim 2$;
 \begin{equation} \Lambda\quad\rightarrow\quad\Lambda_{\rm s}\equiv \Lambda_0 \ {\rm sgn}[z_c-z],
\label{Lambdas}
 \end{equation}
 with $\Lambda_0>0$, and where ${\rm sgn}[x]=-1,0,1$ for $x<0$, $x=0$
 and $x>0$, respectively. It is important to note that besides resolving the $H_0$ and $S_8$ tensions, $\Lambda_s$CDM achieves quite a good fit to Lyman-$\alpha$
 data provided $z_c \lesssim 2.3$~\cite{Akarsu:2019hmw}, and it is in agreement with the otherwise puzzling observations of the James Webb Space Telescope~\cite{Adil:2023ara,Menci:2024rbq}.

At this stage, it is worthwhile to note two caveats of $\Lambda_s$CDM:
\begin{itemize}[noitemsep,topsep=0pt]
\item The analysis in~\cite{Akarsu:2023mfb} is based on the (angular) transversal two dimensional (2D) BAO data on the shell, which are less model dependent than the 3D BAO data. This is because the 3D BAO data sample relies on $\Lambda$CDM to determine the distance to the spherical shell, and hence could potentially introduce a bias when analyzing beyond $\Lambda$CDM models~\cite{Bernui:2023byc, Gomez-Valent:2023uof}. When using  2D BAO data one can accommodate simulateously the actual SH0ES $H_0$ measurement and the angular diameter distance to the last scattering surface, but the effective energy density must be negative for $z \gtrsim 2$.
\item The abrupt behavior of the transition in the vacuum energy density, which is driven by the signum function, leads to a hidden sudden singularity at $z_c$~\cite{Barrow:2004xh}. It is easily seen that the scale factor $a$ of $\Lambda_s$CDM is continuous and non-zero at $t = t_c$, but its first derivative $\dot a$ is discontinuous, and its second derivative $\ddot a$ diverges. However, the sudden singularity yields a minimal impact on the formation and evolution of cosmic bound structures, thereby preserving the viability of $\Lambda_s$CDM~\cite{Paraskevas:2024ytz}.
\end{itemize}

The rapid nature of the sign-switching cosmological constant posits a challenging problem in identifying a concrete theoretical model able to accomodate the transition from an anti-de Sitter  (AdS) to a de Sitter (dS) space. The problem is actually exacerbated by the AdS distance conjecture, which states that 
there is an arbitrarily large distance between AdS and dS vacua in metric space~\cite{Lust:2019zwm}. Having said that, the phenomenological success of $\Lambda_s$CDM, despite its simplistic structure, provides robust motivation to search for possible underlying physical mechanisms that can accommodate the conjectured AdS $\to$ dS transition, and many theoretical realizations are rising to the challenge~\cite{Anchordoqui:2023woo,Akarsu:2024qsi,Akarsu:2024eoo}. In this paper we reexamine the ideas introduced in one of these theoretical realizations~\cite{Anchordoqui:2023woo} to validate its main hypothesis and characterise free parameters of the model.

By combining swampland conjectures with observational data, it was
recently suggested that the cosmological hierarchy problem (i.e. the
smallness of the dark energy in Planck units) could be understood as
an asymptotic limit in field space, corresponding to a
decompactification of one extra (dark) dimension of a size in the
micron range~\cite{Montero:2022prj}. More recently, it was shown that the Casimir forces of fields inhabiting this dark dimension could drive an AdS $\to$ dS transition in the vacuum energy~\cite{Anchordoqui:2023woo}.  Within this set up the Standard
Model (SM) is localized on a D-brane transverse to the compact fifth
dimension, whereas gravity spills into the dark dimension~\cite{Antoniadis:1998ig}. The five-dimensional (5-dim) Einstein-de Sitter gravity action
\begin{equation}
S_{5}=\int[d^4x\, dy] \left({\frac{1}{2}}M_*^{3}{\cal R}^{(5)}-\Lambda_{5}\right)
\end{equation}
reduced to four dimensions in a circle
\begin{equation}
S_{4}=\int[d^4x] \left({\frac{1}{2}}M_p^{2}{\cal R}^{(4)}-{\frac{3}{4}}M_p^2\left({\frac{\partial R}{R}}\right)^2
-(2\pi \langle R \rangle)^{2}\frac{\Lambda_{5}}{2\pi R}\right)
\end{equation}
has a runaway potential inherited from the 5-dim cosmological term
\begin{equation}
V_{\Lambda_5}(R) ={\frac{M_p^2}{M_*^{3}}}{\frac{\Lambda_{5} \  \langle R \rangle}{R}} \,,
\end{equation}
where ${\cal R}^{(d)}$ is the $d$-dimensional curvature scalar,
$M_*$ is the species scale, $\Lambda_{5}$ is a positive 5-dim cosmological constant,
$\langle R\rangle$ is the vacuum expectation value of the modulus controlling the 
radius (or radion field) $R$, 
and 
\begin{equation}
M_p = (M_*^3 \ 2\pi \ \langle R \rangle)^{1/2} 
\end{equation}
is the 4-dim reduced Planck mass~\cite{Anchordoqui:2023etp}. For notational simplicity, we use brackets in the measure transforming as a density under 5-dim diffeomorphisms. Now, if the
5-dim cosmological constant is small, then the quantum contribution of
the lightest 5-dim modes  to the effective potential becomes
relevant. Their contribution at one-loop level can be identified with
the Casimir energy~\cite{Arkani-Hamed:2007ryu}. Bearing this in mind, the 
effective 4-dim potential of the radion can be written as  
\begin{equation}
  V_{\mathrm{eff}}(R) = \frac{2 \pi \ \Lambda_5 \   \langle R \rangle^2}{R} +\left(\frac{\langle R \rangle}{R} \right)^2 \ T_4  + V_C (R) \,,
\label{V}
\end{equation}
where the second term arises from localised orientifolds and D3-branes  (note that orientifolds and D-branes can be on top of each other) with
total tension $T_4$, and 
\begin{equation}
V_C (R) =  \sum_i \frac{\pi \ \langle R \rangle^2}{32
   \pi^7 R^6} \ (N_F - N_B) \ \Theta (R_i -R) \,,
 \label{xxx}
\end{equation} 
stands for the quantum corrections to the vacuum energy due to Casimir forces, with $m_i =
 R_i^{-1}$ the masses of the 5-dim fields, $\Theta$ a step function, and $N_F - N_B$  the
 difference between the number of light fermionic and bosonic degrees
 of freedom. The value and in particular the sign of the the uplifted vacuum energy $V_{\mathrm{eff}}(R)$ now depends on the difference $N_F - N_B$~\cite{Anchordoqui:2023woo}. 

\begin{figure}[htb!]
\centering
\includegraphics[scale=0.6]{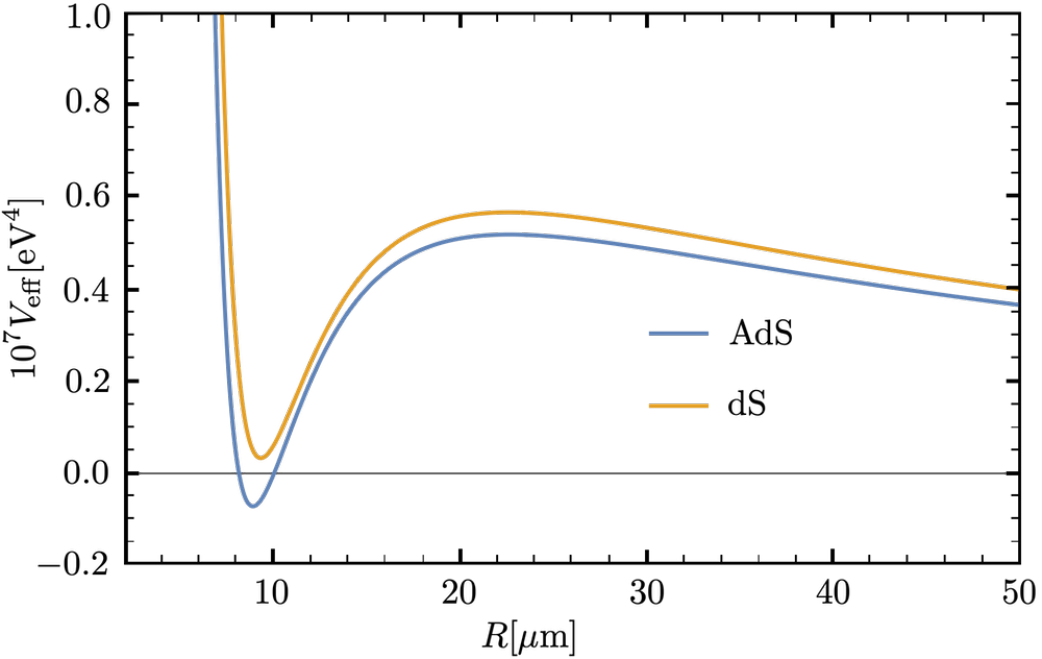} 
\caption{The potential $V_{\mathrm{eff}}(R)$  for $(\Lambda_5)^{1/5} =
    22.6~{\rm meV}$ and $|T_4|^{1/4} =24.2~{\rm meV}$, considering $N_F-N_B = 6$
    (AdS) and $N_F-N_B = 7$ (dS). From Ref.~\cite{Anchordoqui:2023woo}. \label{figurauno}}
\end{figure}

A minimal set up that accommodates the  AdS $\to$ dS transition
requires a 5-dim mass spectrum containing: the graviton, three
generations of light right-handed neutrinos, and a light real scalar
field $\phi$. In~\cite{Anchordoqui:2023woo} it was hypothesised that the real scalar has a potential
$V(\phi)$ holding two local minima with very small difference in vacuum energy
and bigger curvature (mass) of the lower
one, such that at $z_c \sim 2$ the
false vacuum could ``tunnel'' to its true vacuum state. After the
quantum tunneling $\phi$ becomes more massive and its contribution
to the Casimir energy becomes exponentially suppressed. The tunneling
process then  changes the difference between the total number of
fermionic $N_F$ and bosonic $N_B$ degrees of freedom contributing to
the quantum corrections of the vacuum energy. As shown in Fig.~\ref{figurauno}, the change of $N_F-N_B$ could
drive the required AdS $\to$ dS transition at $z_c \sim 2$ (keeping the size of the compact space approximately fixed), if $\Lambda_5^{1/5} \sim 22.6~{\rm meV}$, $|T_4|^{1/4} \sim 24.2~{\rm meV}$, and $\langle R \rangle \sim 10~\mu{\rm m}$.\footnote{We take $h = c = 1$ and so $1~\mu{\rm m} \simeq 0.806~{\rm eV}^{-1}$.}

More concretely, the compactification radius changes from $R_0 \simeq 9.0~\mu{\rm m}$ (in the AdS phase)  to $R_0 \simeq 9.4~\mu{\rm m}$ (in the dS phase). This corresponds to a variation in Newton's gravitational constant of
\begin{equation}
   \frac{\Delta G_N}{G_N} = \frac{\Delta R_0}{R_0} \simeq 0.04 \, .
 \label{deltaG}
 \end{equation}  
 A plethora of investigations have set constraints on the variation of $\Delta G_N/G_N$ over cosmic timescales. An early study, based on the abundances of light elements
 (i.e. D, {$^3$He}, {$^4$He}, and {$^7$Li}) yielded $-0.3 <
\Delta G_N/G_N < 0.4$~\cite{Accetta:1990au}. This study was carried out on the assumption that other
physical constants remained fixed. Subsequently, a study which incorporated constraints from extra relativistic degrees of
freedom  translated the limits on the speed-up factor of
the early universe into bounds on $\Delta G_N/G_N$, leading to
$-0.10 < \Delta G/G_N < 0.13$~\cite{Cyburt:2004yc}. More recent work 
refined this range to $-0.040 < \Delta G_N/G_N < 0.006$, using
improved observational data and updated cosmological
assumptions~\cite{Yeh:2022heq}. It is important to stress that the bound reported in~\cite{Yeh:2022heq} follows from $\Delta G_N/G_N = (7/43) (N_{\rm eff} -3)$, using an effective number of neutrino-like species $N_{\rm eff} = 2.898 \pm 0.141$. Should we use instead the 95\%CL bound reported by the Planck Collaboration $N_{\rm eff} <3.34$~\cite{Planck:2018vyg}, the upper limit on  $\Delta G_N/G_N < 0.05$ is consistent with (\ref{deltaG}).  Using the PRIMAT code and assuming
variations only in $G$ while keeping all other cosmological parameters fixed leads to: $\Delta G_N/G_N = 0.01^{+0.06}_{-0.05}$~\cite{Alvey:2019ctk}. In addition, WMAP and SDSS data constrain the total variation of $\Delta G_N/G_N$ since the epoch of recombination as $-0.083 <\Delta G_N/G_N < 0.095$ at $2\sigma$CL~\cite{Wu:2009zb}. Last but not least, combining information from Planck 2018 CMB
temperature, polarization, and lensing, together with a compilation of BAO measurements
from the Baryon Oscillation Spectroscopic Survey (BOSS), leads to $-0.13 < \Delta G_N/G_N  < 0.049$ at $2\sigma$CL~\cite{Ballardini:2021evv}. Thus, we conclude that the change in radius is small enough so that there is no conflict with observations. Actually, the variation of $\Delta G_N/G_N$ is within the observationally motivated range (estimated in~\cite{Efstratiou:2025xou} to be between 1\% and 7\%).

In this paper we particularise the study carried out in~\cite{Antoniadis:2024ent} to determine the shape of the potential $V (\phi)$ that would allow $\phi$ to undergo the hypothesised Coleman-de Luccia (CdL) transition~\cite{Coleman:1980aw} and at the same time would exponentially suppress the contribution of the scalar field to $V_C(R)$ for $z \lesssim 2$.\footnote{We note in passing the hypothesised CdL transition is from dS to Minkowski space.} The layout of the paper is as follows.
In Sec.~\ref{sec:2}, we briefly review the basics of the bounce configuration and their relation to the second functional variation of the Euclidean action, which has a negative mode~\cite{Callan:1977pt}. The existence and uniqueness of such a negative mode is a criterion for the bounce to mediate vacuum transitions. We thus  investigate the interplay of negative and Kaluza-Klein (KK) modes. We demand that the system has a single negative mode and therefore a KK tower with positive modes. This condition leads to a stability constraint on the radius of the compact dimension. In Secs.~\ref{sec:3} and \ref{sec:4}, we set up conditions on the shape of the potential $V(\phi)$ under which a late time AdS $\to$ dS transition could be driven by Casimir forces. In Sec.~\ref{sec:5} we equate the required lifetime of the false vacuum to the CdL (semi-classical) tunneling rate per unit volume to obtain an estimate of the on-shell Euclidean action of the bounce. The normalization factor, which is a one-loop determinant, is approximated using dimensional analysis. Because of the exponential dependence of the vacuum decay rate on the bounce factor, this approximation does not lead to large errors in our calculations. Adjusting the free parameters of the model to fiducial values we study two possible regimes of the bounce factor. We show that to accommodate the required transition at $z_c$ the size of the CdL instanton must be larger than the compactification radius. The paper wraps up in Sec.~\ref{sec:6} with some conclusions. 

Before proceeding, we pause to note that
the 5-dim fields characterising the deep infrared region of the dark sector contribute to the effective number of relativistic neutrino-like species. We also note that the addition of extra relativistic degrees of freedom  does not spoil the $\Lambda_s$CDM resolution of the $H_0$ and $S_8$ tensions~\cite{Anchordoqui:2024gfa,Yadav:2024duq}. 

\section{Bounce, negative mode and Kaluza-Klein tower}
\label{sec:2}

It has long been known that quantum tunnelling causes vacuum decay, both with and without gravity~\cite{Coleman:1977py,Callan:1977pt,Coleman:1980aw}. Broadly speaking, the decay of a metastable vacuum embodies a first-order phase transition, in which  the order parameter changes from a metastable phase to a stable phase. The transition could materialize via  nucleation of bubbles produced by quantum or statistical fluctuations. Bubbles below a critical size collapse because of the overwhelming surface energy cost, whereas the larger bubbles grow rapidly and after some time fill space to complete the phase transition.

The classical technique for computing the bubble nucleation rate in field theory requires solving the equations of motion (EoM) in Euclidean signature to find the saddle point of the path integral (a.k.a. the bounce).
The tunnelling probability per unit volume is given by
\begin{equation}
\frac{\Gamma}{V_{D-1}} = A e^{-B/\hslash} \ \left[ 1 + {\cal O} (\hslash) \right] \,,
\label{uno}
\end{equation}
where $D$ is the number of moduli of the bounce solution to the EoM that makes the main contribution to the vacuum decay, as we will explain in detail later. $V_{D-1}$ is the volume of $(D-1)$-dimensional Euclidean space $\bfR^{D-1}$, which is the volume of the spatial subspace of the moduli space corresponding to the degrees of freedom of the spatial translation of the bounce, $B$ is the on-shell Euclidean action at the bounce,
and $A$ is a normalization factor that consists of one-loop determinants~\cite{Coleman:1980aw}. Since the domain of integration of the on-shell Euclidean action $S_{\rm E}[\phi]$ naturally splits into three regions, following~\cite{Coleman:1980aw} we identify three contributions to the total bounce factor  
\begin{equation}
    B \equiv S_{\rm E} [\phi_\rB] - S_{\rm E} [v_+] = B_{\rm in} + B_{\rm wall} + B_{\rm out} \,,
\end{equation}
where $\phi_\rB$ denotes the bounce solution, and $v_+$ is the expectation value of $\phi$ at the false vacuum. We will also use $v_-$ as the expectation value of $\phi$ at the true vacuum. These regions are defined in terms of the interior, the thin wall, and the exterior of the bubble. The false vacuum  $v_+$ resides outside the bubble and the true vacuum $v_-$ resides inside. We can distuinguish two distinct regimes for the bounce configuration that are characterised by a 4-dim space with a compact extra dimension (in which $D=4$) 
and a 5-dim non-compact space (in which $D=5$). 

Now, linear fluctuations around a stationary point of the action describing vacuum decay must allow only one negative mode.  This is because the decay rate of a metastable vacuum is determined by the imaginary part of the energy as computed by the effective action~\cite{Callan:1977pt}, and hence only solutions that contribute an imaginary part to the vacuum energy will contribute to metastability. In the reminder of this section, we discuss the interplay of negative and KK modes. To identify the 4-dim non-compact negative mode we will follow the argument given in~\cite{Lee:2014uza}, which in our 5-dim case with a compact fifth dimension will be dressed by a KK tower. We demand that the system has a single negative mode and
therefore a KK tower on top of it only with positive modes. This leads to a constraint on the radius of the compact dimension, which is translated into a critical value above which the bounce is unstable and fails to describe the vacuum decay.

In the following, we will consider a simple setup for the vacuum decay, which consists of a real scalar field $\phi$ in a potential $V$ in a $d$-dimensional spacetime. Its Euclidean action $S_\rE[\phi]$ is given by
\begin{align}
S_\rE[\phi]=\int\!d\tau d^{d-2}xdy\,\left[\frac{1}{2}(\pd\phi)^2+V(\phi)\right] \, ,
\end{align}
where $\tau$ is the Euclidean time, $d^{d-2}x=dx^1 \cdots dx^{d-2}$ with coordinates $x^1,\cdots,x^{d-2}$ for the $(d-2)$-dimensional flat space $\bfR^{d-2}$, and $y$ refers to a coordinate for the $d$-th dimension, which is either a real line $\bfR^1$ or a line interval $S^1/\mathbf{Z}_2$ with two ends of length $\pi R_0$.

We will work in the gravity decoupling limit $\kappa_d\to0$, which will be justified a posteriori in Sec.~\ref{sec:5}. In the absense of gravity we can shift the vacuum energy arbitrarily and for convenience we choose to normalise it to zero in the true vacuum.

\subsection{Non-compact $d$-dimensions}  
We start from the case with the $d$-dimensional non-compact spacetime in the absence of gravity, thereby $x^M=(\tau,x^1,\cdots,x^{d-2},y)$ forming coordinates of $d$-dimensional Euclidean flat spacetime $\bfR^d$. The EoM from the action $S_\rE[\phi]$ reads
\begin{equation}\label{phi-EOM}
-\triangle_d\phi+V'(\phi) = 0 \, ,
\end{equation}
where $\triangle_d$ is the $d$-dimensional Laplacian $\pd^M\pd_M$.
Since its bounce solution $\phi_\rB$ is spherical, namely $O(d)$-symmetric, it is convenient to work with the spherical coordinates. The translation symmetry of the system allows free choice of the position of the centre of the bounce. Therefore, we may define the radial coordinate by 
\begin{align}
\xi=\sqrt{(x-x_0)^M(x-x_0)_M} \, , 
\end{align}
where $x_0^M$ refers to the position of the centre of the bounce in $\bfR^d$, playing the role of the moduli of the bounce. Namely, the moduli space of the bounce in the present case is $\bfR^d$. 

In terms of spherical coordinates of $\bfR^d$, the bounce is just a function of $\xi$ and satisfies the EoM
\begin{equation}\label{phi-EOM-bounce}
\ddot{\phi}_\rB(\xi)+\frac{d-1}{\xi} \dot{\phi}_\rB(\xi) = V'(\phi_\rB) \, ,
\end{equation}
where each dot means $d/d\xi$.

\paragraph{Linear fluctuations}
Eigenmodes of linear fluctuations around the bounce $\phi_\rB$ satisfy the equation
\begin{equation}\label{eigeneq-bounce}
\left[ -\triangle_d+V''(\phi_{\rB}) \right] \eta = h \eta \, ,
\end{equation}
where $\eta$ is the eigenmode with eigenvalue $h$ satisfying $\eta=0$ at $\xi=0$ and $\xi\to\infty$. In spherical coordinates the Laplacian reads 
\begin{equation}\label{Laplaciand}
\triangle_d = \frac{\partial^2}{\partial \xi^2} + \frac{d-1}{\xi}\frac{\partial}{\partial \xi}+ \frac{1}{\xi^2} \triangle_{S^{d-1}} \, ,
\end{equation}
where $\triangle_{S^{d-1}}$ is the Laplacian on the unit sphere $S^{d-1}$ with respect to its angular coordinates, playing the role of the quadratic Casimir of $O(d)$. Let us summarise its properties in a bit more detail.\footnote{For more details about spherical harmonics in general dimensions, see, for example, Section 2 of~\cite{Isono:2019wex}.} Spherical harmonics of spin $\ell$ on $S^{d-1}$ form a basis of the spin-$\ell$ representation space of $O(d)$ of dimension $\mathfrak{d}_{\ell,d}$ given by
\begin{align}
\mathfrak{d}_{\ell,d}=\frac{(\ell+d-3)!(2\ell+d-2)}{\ell!(d-2)!} \,,
\end{align}
and the spherical Laplacian $\triangle_{S^{d-1}}$ has eigenvalue $\ell(\ell+d-2)$ in the spin-$\ell$ representation $O(d)$. Let us denote spherical harmonics in the spin $\ell$ representation of $O(d)$ by $Y^d_{\ell,m}(\Om)$, where $m=1,2,\cdots,\mathfrak{d}_{\ell,d}$ and $\Om$ collectively denotes angular coordinates of $S^{d-1}$. We can then expand any normalisable function $f$ on $S^{d-1}$ as $f=\sum_{\ell=0}^\infty\sum_{m=1}^{\mathfrak{d}_{\ell,d}}c_{\ell,m}Y^d_{\ell,m}$ with coefficients $c_{\ell,m}$, and the Laplacian $\triangle_{S^{d-1}}$ on $S^{d-1}$ satisfies
\begin{align}
\triangle_{S^{d-1}}Y^d_{\ell,m}=\ell(\ell+d-2)Y^d_{\ell,m} \,.
\end{align}
Therefore, each eigenfunction of the eigenmode equation \eqref{eigeneq-bounce} can be written in the form:
\begin{align}\label{eigenfunc-O(d)}
\psi_{\ell}(\xi)Y^d_{\ell,m}(\Om) \,, \quad
\ell\geq 0 \,, \quad 1\leq m\leq\mathfrak{d}_{\ell,d} \,,
\end{align}
where the radial part $\psi_{\ell}(\xi)$ satisfies the \textit{radial} eigenmode equation with an eigenvalue $h_\ell$,
\begin{align}
&\tilde O_{\rB,\ell}\,\psi_{\ell}=h_\ell\,\psi_{\ell} \,, \quad
\psi_\ell(0)=\psi_\ell(\infty)=0 \,, \label{radeigenmodeeq-d1} \\
&\tilde O_{\rB,\ell}:=-\frac{\partial^2}{\partial \xi^2}-\frac{d-1}{\xi}\frac{\partial}{\partial \xi}-\frac{\ell(\ell+d-2)}{\xi^2}+V''(\phi_{\rB}) \,, \label{radeigenmodeeq-d2}
\end{align}
so that $\psi_\ell(\xi) Y^d_{\ell,m}(\Om)$ corresponds to eigenvalue $h=h_\ell$. Note that it is independent of the label $m$ of spherical harmonics for each spin $\ell$ and hence all $\psi_\ell(\xi) Y^d_{\ell,m}(\Om)$ $(1\leq m\leq\mathfrak{d}_{\ell,d})$ have the same eigenvalue $h_\ell$.

In summary, denoting the set of all eigenvalues of $\tilde O_{\rB,\ell}$ by $\sigma(\tilde O_{\rB,\ell})$, we can obtain all eigenvalues of the eigenmode equation \eqref{eigeneq-bounce} by collecting $\sigma(\tilde O_{\rB,\ell})$ for all spins $\ell$; namely $\cup_{\ell \geq 0}\,\sigma(\tilde O_{\rB,\ell})$.
Each eigenvalue $h_\ell$ corresponds to the eigenfunctions of the form in \eqref{eigenfunc-O(d)} and has multiplicity $\mathfrak{d}_{\ell,d}$.

\paragraph{Negative mode}
As mentioned above, the eigenmode equation \eqref{eigeneq-bounce} yields one negative eigenmode.
We can demonstrate it explicitly in the thin-wall approximation starting from the EoM of the bounce \eqref{phi-EOM-bounce}, following~\cite{Lee:2014uza}.
Differentiating \eqref{phi-EOM-bounce} with respect to $\xi$ gives
\begin{equation}\label{eigenequation}
\left[ \frac{d^2}{d\xi^2} + \frac{d-1}{\xi} \frac{d}{d\xi} - V''(\phi_\rB) \right] \dot\phi_\rB = \frac{d-1}{\xi^2} \dot\phi_\rB \, .
\end{equation}
This is not in the form of the eigenmode equation \eqref{eigeneq-bounce} because its right hand side depends on $\xi$. 
However, in the thin-wall approximation~\cite{Coleman:1977py,Coleman:1980aw}, the bounce solution exhibits non-trivial dependence on $\xi$ only on the wall, which is located at $\xi=\bar\xi$, having a constant profile in- and outside the wall.\footnote{See also Appendix~\ref{app:B} for the thin-wall approximation.} This implies that $\dot\phi_\rB$ has a sharp peak around the wall $\xi=\bar\xi$ and is zero in- and outside the wall. Therefore, Eq.~\eqref{eigenequation} can be rewritten as
\begin{equation}
\left[ -\frac{d^2}{d\xi^2} - \frac{d-1}{\xi} \frac{d}{d\xi} + V''(\phi_\rB) \right]\dot\phi_\rB = -\frac{d-1}{\bar{\xi}^2}\dot\phi_\rB \, .
\label{rewritten}
\end{equation}
Note that $\dot\phi_\rB$ satisfies $\triangle_{S^{d-1}}\dot\phi_\rB=0$ because $\phi_\rB$ is independent of the angular coordinates of $S^{d-1}$. Therefore, comparing Eq.~\eqref{rewritten} with the eigenmode equation \eqref{radeigenmodeeq-d1}, we find that Eq.~\eqref{rewritten} is the eigenmode equation of spin 0 with the eigenfunction $\dot\phi_\rB$ corresponding to a negative eigenvalue $h_0$ given by\footnote{Note that spin 0 has only one spherical harmonic, which is just a constant.}
\begin{equation}
h_0 = -\frac{d-1}{\bar{\xi}^2} \,,
\end{equation}
where the subscript $0$ refers to the spin. This negative mode reflects the change in the size of the bounce, which is nothing but the instability associated with the vacuum decay.

\paragraph{Zero modes}
The moduli of the bounce solution are associated with zero modes of the linear fluctuations around the bounce. The zero modes are given by eigenfunctions $\pd_M\phi_B$ $(M=1,\cdots,d)$ of \eqref{radeigenmodeeq-d1} for spin 1. Indeed, by acting on the EoM of the bounce \eqref{phi-EOM-bounce} with $\pd_M$, we obtain 
\begin{equation}
\left[ -\frac{\partial^2}{\partial\xi^2} - \frac{d-1}{\xi} \frac{\partial}{\partial\xi} + V''(\phi_\rB) + \frac{d-1}{\xi^2} \right]\pd_M\phi_\rB = 0 \,.
\label{rewritten-zero-d}
\end{equation}
The spin is read off by noticing that $d-1$ is equal to $\ell(\ell+d-2)$ with $\ell=1$. Note that $\pd_M\phi_\rB=(x_M/\xi)\dot\phi_\rB$ and functions $x_M/\xi$ are $d$ spherical harmonics of spin 1. Therefore, the radial operator $\tilde O_{\rB,1}$ has one zero eigenvalue and the number of the zero modes $d$ is nothing but the dimension $\mathfrak{d}_{1,d}$ of the spin 1 representation of $O(d)$.

\subsection{$d$-dimensions with a compact direction} 
\label{subsec:O(4)U(1)}
Let us proceed to the case where the extra ($d$-th) direction is an interval of length $\pi R$ with coordinate $y$, thereby the Euclidean spacetime being $\bfR^{d-1}\times S^1/\mathbf{Z}_2$. We use $x^\mu$ as coordinates of $\bfR^{d-1}$. The equation of motion is
\begin{equation}\label{phi-EOM-S1}
-\triangle_{d-1}\phi-\frac{\pd^2\phi}{\pd y^2}+V'(\phi) = 0 \, .
\end{equation}
where $\triangle_{d-1}$ is the Laplacian $\pd^\mu\pd_\mu$ in $\bfR^{d-1}$ with respect to $x^\mu$. In this case, the bounce is taken to be independent of the $S^1$ coordinate $y$, and has spherical symmetry in the non-compact part $\bfR^{d-1}$; namely the bounce is $O(d-1)\times U(1)$ symmetric~\cite{Antoniadis:2024ent}. The translation symmetry of $\bfR^{d-1}$ allows free choice of the centre of the bounce. Therefore, we may define the radial coordinate as 
\begin{align}
\xi=\sqrt{(x-x_0)^\mu(x-x_0)_\mu} \, ,
\end{align}
where $x_0^\mu$ refers to the centre of the bounce in $\bfR^{d-1}$. On the other hand, the degrees of freedom in choosing the origin of the $S^1$ coordinate does not provide a modulus because the bounce is independent of $y$ and hence translation of the bounce in $S^1$ does nothing, giving the \textit{identical} configuration, while different moduli $x_0^\mu$ give different (translated) configurations. In conclusion, the moduli space of the bounce in the present case is $\bfR^{d-1}$.

In spherical coordinates of $\bfR^{d-1}$, the EoM of the bounce reads
\begin{equation}\label{phi-EOM-bounce-O(d-1)}
\ddot{\phi}_\rB(\xi)+\frac{d-2}{\xi} \dot{\phi}_\rB(\xi) = V'(\phi_\rB) \, ,
\end{equation}
which is identical with the EoM of the bounce in $\bfR^{d-1}$ without the extra dimension.

\paragraph{Linear fluctuations}
The eigenmode equation around the bounce reads
\begin{equation}\label{eigeneq-bounce-O(d-1)}
\left[ -\triangle_d+V''(\phi_{\rB}) \right] \eta = h \eta \,.
\end{equation}
The differential operator $-\triangle_d+V''(\phi_{\rB})$ can be expressed in spherical coordinates as
\begin{equation} \label{eigenmodeeq-O(d-1)}
\left[ -\frac{\partial^2}{\partial \xi^2}-\frac{d-2}{\xi}\frac{\partial}{\partial \xi}-\frac{1}{\xi^2} \triangle_{S^{d-2}}-\frac{\pd^2}{\pd y^2}+V''(\phi_\rB) \right] \eta = h \eta \, .
\end{equation}
where we used the decomposition of the Laplacian \eqref{Laplaciand} with the replacement $d \to d-1$. 
Since $\phi_\rB$ is independent of $y$, this has a separable form.
To find mode functions for fluctuations, we need to specify boundary conditions on $\eta$ at the two endpoints $y=0,\pi R_0$ of the interval $S^1/\mathbf{Z}_2$. In general, $\eta$ can be expanded in $\sin$ and $\cos$ wave functions. Since fluctuations without extra-dimensional dependence are allowed to contribute to the one-loop determinant, we need a KK zero (constant) mode. We therefore adopt $\cos$ wavefunctions satisfying the Neumann boundary conditions. This choice is consistent with the classical bounce configuration since it is independent of $y$ and hence satisfies the Neumann boundary condition trivially. Taking them into account, we can express each eigenfunction in the form:
\begin{align}
&\psi_{\ell}(\xi)Y^{d-1}_{\ell,m}(\Om)\cos(sy/R_0) \,, \label{eigenfunc-O(d-1)} \\
&\ell=0,1,2,\cdots\,, \quad m=1,\cdots,\mathfrak{d}_{\ell,d-1}\,, \quad s=0,1,2,\cdots \, ,
\end{align}
where $\Om$ collectively denotes the angular coordinates of $S^{d-2}$, $R_0$ is the radius of $S^1$, and $s$ labels the KK modes. 
The radial part $\psi_{\ell}$ satisfies the radial eigenmode equation with an eigenvalue $h_\ell$,
\begin{align} 
&\tilde O_{\rB,\ell}\,\psi_{\ell} = h_\ell \,\psi_{\ell} \,, \quad \psi_\ell(0)=\psi_\ell(\infty)=0 \,, \label{eigenpsi-O(d-1)1} \\
&\tilde O_{\rB,\ell} := -\frac{\partial^2}{\partial \xi^2}-\frac{d-2}{\xi}\frac{\partial}{\partial \xi}-\frac{\ell(\ell+d-3)}{\xi^2}+V''(\phi_\rB) \,, \label{eigenpsi-O(d-1)2}
\end{align}
so that the function \eqref{eigenfunc-O(d-1)} corresponds to eigenvalue
\begin{align}
h_\ell+\frac{s^2}{R_0^2} \,.
\end{align}
In other words, given an eigenfunction $\psi_{\ell}(\xi)$ of \eqref{eigenpsi-O(d-1)1} with spin $\ell$, the set 
\begin{align} \label{KKtower}
\{\psi_{\ell}(\xi)Y^{d-1}_{\ell,m}(\Om)\cos(sy/R_0)\}_{s=0,1,2,\cdots}
\end{align}
gives a KK tower of eigenmodes with eigenvalues $\{h_\ell+s^2/R_0^2\}_{s=0,1,2,\cdots}$ for each spherical harmonic $Y^{d-1}_{\ell,m}$ of $O(d-1)$ spin $\ell$.

In summary, the set of all eigenvalues of the eigenmode equation \eqref{eigenmodeeq-O(d-1)} is obtained as follows. We extend the set of all eigenvalues of $\tilde O_{\rB,\ell}$ to the set $\sig_{\mathrm{KK}}(\tilde O_{\rB,\ell})$ by assigning the KK tower of the form $\{h_\ell+s^2/R_0^2\}_{s=0,1,2,\cdots}$ to each eigenvalue $h_\ell$ of $\tilde  O_{\rB,\ell}$, and then collect the sets $\sig_{\mathrm{KK}}(\tilde O_{\rB,\ell})$ for all spins $\ell$; namely $\cup_{\ell\geq 0}\,\sig_{\mathrm{KK}}(\tilde O_{\rB,\ell})$.
Each eigenvalue in $\sig_{\mathrm{KK}}(\tilde O_{\rB,\ell})$ corresponds to the eigenfunctions of the form in \eqref{eigenfunc-O(d-1)} and has multiplicity $\mathfrak{d}_{\ell,d-1}$.

\paragraph{Zero and negative modes, stability condition}
Since the eigenmode equation \eqref{eigenpsi-O(d-1)1} is nothing but \eqref{radeigenmodeeq-d1} with $d$ replaced by $d-1$, it yields one negative mode for spin 0, and $(d-1)$ zero modes for spin 1. 
The zero modes are given by eigenfunctions $\pd_\mu\phi_\rB$ having spin 1 and KK label $s=0$. 
The negative mode is given in the thin-wall approximation by eigenfunction $\dot\phi_\rB$ of spin 0 and KK label $s=0$ and has eigenvalue
\begin{align}
h_0=-\frac{d-2}{\bar\xi^2} \,,
\end{align}
where $\bar\xi$ is the position of the wall of the bounce with $O(d-1)\times U(1)$ symmetry. Therefore, we have the KK tower on top of it with eigenvalues
\begin{align} 
h^{(s)}_0=-\frac{d-2}{\bar\xi^2}+\frac{s^2}{R_0^2} \, .
\end{align}
Now, we require that \textit{the system should have only one negative mode $\psi_0$}. Otherwise, the system would have instabilities other than the vacuum decay and hence the bounce configuration would fail to describe the decay. This requirement forces the KK excitations to have positive eigenvalues, $h_0^{(s)}>0$ for $s>0$, yielding the following upper bound on the compactification radius:
\begin{align}
R_0^2<\frac{\bar\xi^2}{d-2} \, .
\end{align}
Since we will work on the five-dimensional case $d=5$, we present the stability condition in this particular case,
\begin{equation}
R_{0}^2 < \frac{\bar{\xi^2}}{3} \, .
\label{ultima}
\end{equation}
This stability condition based on the negative mode gives an answer to the puzzle raised in~\cite{Antoniadis:2024ent} about the validity and the transition of the $O(4)\times U(1)$ symmetric bounce to the $O(5)$ symmetric one.

\section{Incompatibility of quartic potentials with hypotheses of the transition}
\label{sec:3}

In this and next sections, we set up concrete scalar potentials and consider constraints on them from the late-time AdS $\to$ dS transition. This section will consider quartic potentials and show that they are incompatible with the condition on the mass hierarchy between the false and true vacua needed for the transition. 

We start from the requirement that the quartic scalar potential $V(\phi)$ in 5-dim should have two minima at $v_+$ and $v_-$. Using the degrees of freedom of shifting $\phi$ and adding a constant to the potential, we are left with three real parameters out of five in a quartic potential. The requirement for the two minima is then sufficient to fix $V'(\phi)$ to
\begin{align}
V'(\phi)=\lm\phi(\phi-v_+)(\phi-v_-) \, , 
\end{align}
where $\lm$ is a positive parameter. We fixed the local maximum between the two minima to $\phi=0$, yielding the constraint $v_+v_-<0$. Integrating $V'$ over $\phi$ and imposing $V(v_-)=0$, we obtain the potential,
\begin{align}\label{phi4_V}
V(\phi)=\frac{1}{4}\lm(\phi-v_-)^2\left[\phi^2+\frac{2}{3}(v_--2v_+)\phi+v_-(v_--2v_+)\right]  \, .
\end{align}
The masses at the minima are given by
\begin{align}
m_\pm^2:=V''(v_\pm)=\lm v_\pm(v_\pm-v_\mp) \, ,
\end{align}
and the difference of the vacuum energies at the minima is 
\begin{align}
\vep:=V(v_+)-V(v_-)=\frac{\lm}{12}(v_--v_+)^3(v_-+v_+) \, , 
\end{align}
which we assume to be positive since $v_+$ is the false vacuum and $v_-$ is the true vacuum. 

\begin{figure}[htb!]
\centering
\includegraphics[scale=0.45]{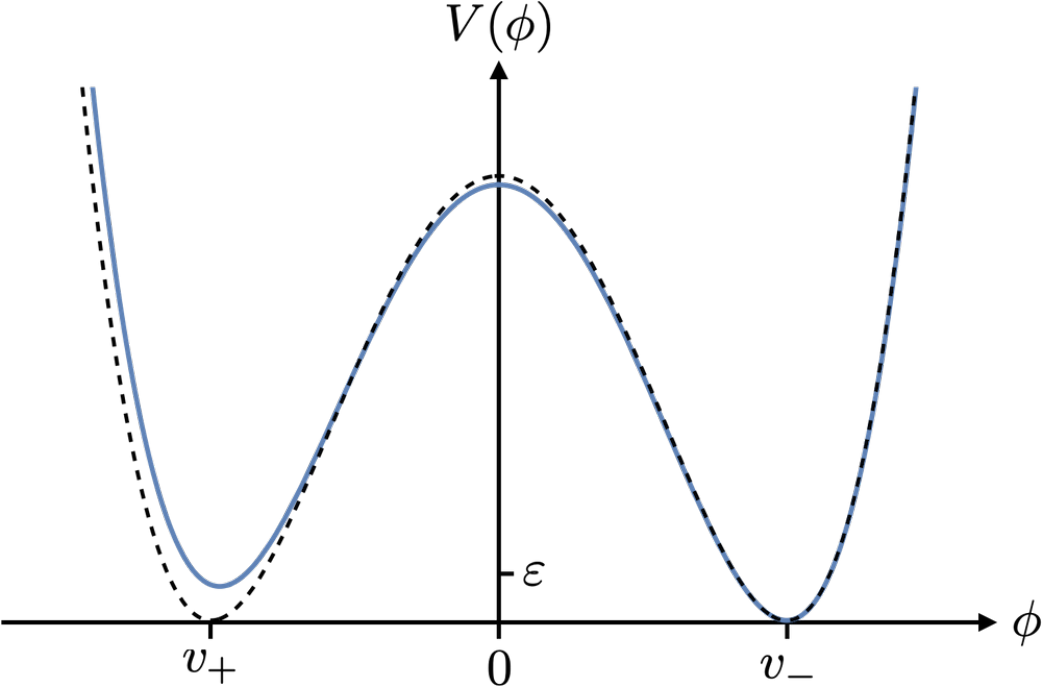} 
\caption{Schematic form of the scalar potential in \eqref{phi4_V}. The dashed black curve represents the $\varepsilon =0$ case while the solid blue curve shows the $\varepsilon \neq 0$ case.
\label{fig:1}}
\end{figure}

As we already mentioned in Section~\ref{sec:2}, we adopt the thin-wall approximation \cite{Coleman:1977py,Coleman:1980aw}, which is justified when the energy difference $\vep$ is sufficiently small, making the thickness of the bounce much smaller than the size of the wall (see Appendix~\ref{app:B}). This condition is translated into our parameters as
\begin{align}
\frac{v_-}{v_+} \simeq -1 \, , 
\end{align}
which makes the mass ratio $m_-/m_+$ almost one,
\begin{align}
\frac{m_-^2}{m_+^2} \simeq 1 \, .
\label{massratio1}
\end{align} 
Thus in this limit, the potential is reduced to the standard double-well quartic form symmetric under $\phi\to-\phi$ (see Fig.~\ref{fig:1}).

On the other hand, let us recall the conditions on the masses $m_\pm$ at the two vacua. For large $m_-$, the Casimir energy density is exponentially suppressed~\cite{Anchordoqui:2023woo}. This implies that the transition from AdS to dS require that the mass at the true vacuum should be approximately one order of magnitude bigger than that at the false vacuum $m_-/m_+ \gg 1$. However, this contradicts the condition on the mass ratio \eqref{massratio1} from the thin-wall approximation. We therefore conclude that quartic potentials are incompatible with the conditions for the AdS $\to$ dS transition within the thin-wall approximation.

The origin of this incompatibility is that in the thin-wall limit, the two vacua have not only equal energy but also equal mass.  This motivates us to start from the potential in the thin-wall limit with equal vacuum energy but different masses at the two vacua.

\section{Sixth order potential}
\label{sec:4}

As suggested at the end of the last section, the arguments for excluding the quartic potentials suggest that the double-well potential should have different masses \textit{already in the thin-wall limit where the two minima have the same energy}. In the following we construct a set of sixth order potentials that fulfill this requirement. Let $V_0$ be the potential in the thin-wall limit we are going to construct. 

We start from making an ansatz on the first derivative of $V_0$. We may reduce seven real parameters in a sixth order potential to five by using the degrees of freedom of shifting $\phi$ and adding a constant to $V_0$. Requiring that $V_0$ should have only three extrema consisting of two minima and one local maximum inbetween, we reach the following ansatz on $V_0'$ with five parameters:
\begin{align}
V_0'(\phi)=\lambda\phi(\phi-v_-)(\phi-v_+)[(\phi-v)^2+M^2] \, ,
\end{align}
where $M$ is a real parameter and $\lambda>0$, and we fixed the local maximum to $\phi=0$. The positive definite factor $(\phi-v)^2+M^2$ ensures that $V_0$ has only three extrema. The masses at the extrema are
\begin{align}
m_\pm^2=V_0''(v_\pm)&=\lambda v_\pm(v_\pm-v_\mp)[(v_\pm-v)^2+M^2]
\end{align}
and
\begin{align}
V_0''(0)&=\lambda v_+v_-(v^2+M^2) \, .
\end{align}
We require that the potential should have two minima at $\phi=v_\pm$ and one maximum at $\phi=0$; namely $V_0''(v_\pm)>0$ and $V_0''(0)<0$, which gives
\begin{align}
v_+v_-<0 \, .    
\end{align}
We assume that the extrema satisfy  $v_+<0<v_-$.

Another requirement is that the two minima have the same potential energy,
\begin{align}
  V_0(v_+)=V_0(v_-) \, ,
\end{align}
which is equivalent to
\begin{align}
\int_{v_+}^{v_-}V_0'(\phi) \ d\phi = 0 \, .
\end{align}
This condition fixes $M^2$ as a function of $v_\pm$ and $v$,
\begin{align}
M^2=\frac{1}{5}\left(-5v^2-2v_+^2-v_+v_--2v_-^2+2v\frac{3v_+^2+4v_+v_-+3v_-^2}{v_++v_-}\right)\,.
\label{47}
\end{align}
The value $V_0(v_\pm)$ at the minima can be tuned by adding a constant to the potential.

To facilitate the analysis of $M^2$, we parameterise $v$ as
\begin{align}
v=\al(v_++v_-) \,,
\label{48}
\end{align}
and introduce
\begin{align}
r:=\frac{v_-}{v_+}<0 \quad {\rm and} \quad
\bt^2:=\frac{M^2}{v_+^2}\,,
\label{49}
\end{align}
so that $\al,r$ and $\bt$ are dimensionless parameters.
Substituting (\ref{48}) and (\ref{49}) into (\ref{47}) we obtain
\begin{align}
5\bt^2=-(5\al^2-6\al+2)r^2-(10\al^2-8\al+1)r-(5\al^2-6\al+2)\,.
\end{align}
Note that $5\al^2-6\al+2>0$ for any real $\al$. The discriminant with $r$ considered a variable reads
\begin{align}
\Delta_r=5(4\al-3)(2\al-1)^2\,.
\end{align}
The positivity $\bt^2>0$ is then satisfied if
\begin{align}
\alpha>\frac{3}{4} ~~ \text{and} ~~ r_+<r<r_-\,,
\end{align}
where the roots $r_\pm$ are defined by
\begin{align}
r_\pm=-\frac{10\al^2-8\al+1\pm(2\al-1)\sqrt{5(4\al-3)}}{10\al^2-12\al+4}\,,    
\end{align}
and are both negative for $\al>3/4$.

\begin{figure}[htb!]
\centering
\includegraphics[scale=0.45]{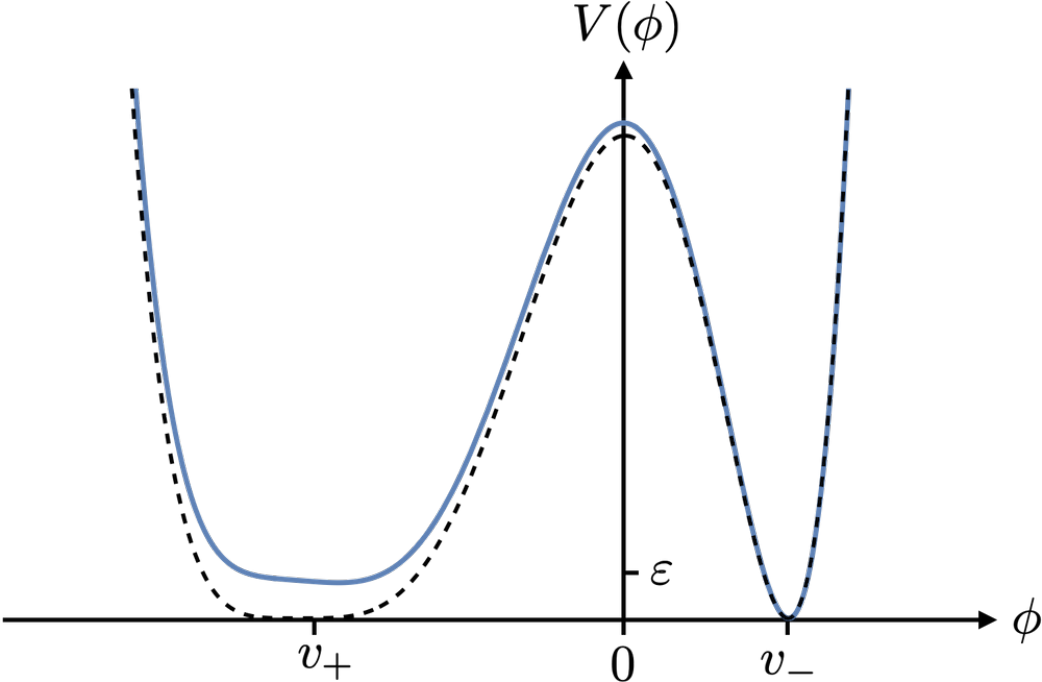} 
\caption{Schematic form of the $\phi^6$ scalar potential in \eqref{phi6_V}. The dashed black curve represents the $\varepsilon =0$ case while the solid blue curve shows the $\varepsilon \neq 0$ case.  }
\label{fig:2}
\end{figure}

Since $m_\pm^2:=V_0''(v_\pm)$, the mass ratio can be written as
\begin{align}
\frac{m_-^2}{m_+^2}=-\frac{v_-}{v_+}\frac{(v_--v)^2+M^2}{(v_+-v)^2+M^2}=-r\frac{[r-\al(1+r)]^2+\bt^2}{[1-\al(1+r)]^2+\bt^2}\,.
\end{align}
Note that $m_+^2=0$ for
\begin{align}
(\al,r)=\left(2,-\frac{1}{2}\right)\,,
\end{align}
pushing $m_-^2/m_+^2 \to \infty$. Note also that $(r_-,r_+)=(-1/2,-2)$ for $\al=2$. This indicates that $m_-^2/m_+^2 \gg 1$ is attainable if $\al=2$ while perturbing $r$ around $-1/2$, i.e.,
\begin{align}\label{alr}
\al=2\,, \quad r=-\frac{1}{2}-\de\,,  
\end{align}
where $\de$ is taken to be a positive, small parameter. 
Using this parametrisation, we find the following relations
\begin{align}
&\frac{v_-}{v_+}=-\frac{1}{2}-\de+\mathcal{O}(\de^2)\,, \quad
v=v_+(1-2\de)+\mathcal{O}(\de^2)\,, \label{parameters-in-delta}\\
&\bt^2=3\de+\mathcal{O}(\de^2)\,, \quad
\frac{m_-^2}{m_+^2}=\frac{3}{8\de}[1+\mathcal{O}(\de)]\,, \quad 
m_+^2=\frac{9}{2}\lm v_+^4\de+\mathcal{O}(\de^2)\,, \label{parameters-in-delta2}
\end{align}
and
\begin{align}\label{V0''(0)}
  V_0''(0)=\frac{1}{2}\lm v_+^4+\mathcal{O}(\de)=-\frac{8}{27}m_-^2+\mathcal{O}(\de) \, .
\end{align}
Under this parametrisation, the first derivative of the potential reads
\begin{align}
V_0'(\phi)=\lm\phi(\phi-v_+)^3(\phi-v_-)+\mathcal{O}(\de)\,,
\end{align}
and so the potential is found to be
\begin{align}\label{V0explicit}
V_0(\phi)=\frac{\lm}{6}(\phi-v_+)^4(\phi-v_-)^2+\mathcal{O}(\de)\,,
\end{align}
where we normalised the vacuum energy to zero.
Once we obtain the double-well potential $V_0$ in the thin-wall limit, the potential for the tunneling can be parametrised as \cite{Coleman:1977py}
\begin{align}
V(\phi)=V_0(\phi)+\varepsilon\frac{\phi-v_-}{v_+-v_-}\,,
\label{phi6_V}
\end{align}
so that the difference of the vacuum energies is still $\vep$.\footnote{Exactly speaking, the vacuum expectation values of $\phi$ at the two vacua $\hat v_\pm$ of the tunneling potential $V$ are different from $v_\pm$: $\hat v_+=v_++\mathcal{O}(\vep^{1/3})$ and $\hat v_-=v_-+\mathcal{O}(\vep)$. The vacuum energies then change into $V(\hat v_+)=\vep+\mathcal{O}(\vep^{4/3})$ and $V(\hat v_-)=\mathcal{O}(\vep^2)$. Therefore, the difference of the vacuum energies is $\vep+\mathcal{O}(\vep^{4/3})$.} A schematic representation of the potential is shown in Fig.~\ref{fig:2}.\footnote{In this figure, the mass ratio $m_-/m_+$ was set to a number much larger than 1 for visualisation. However, as we will see in the next section, it is sufficient to assume $m_-/m_+\gtrsim5$ for the AdS $\to$ dS transition.}

Bearing this in mind, the difference of on-shell Euclidean actions is largely simplified and can be computed explicitly,
\begin{align}
S_{1}:=  \int_{v_{+}}^{v_{-}} \! \sqrt{ 2\left[ V_{0}(\phi)-V_{0}(v_{+}) \right]} \, d\phi  \simeq\frac{9\sqrt{3}}{64} \sqrt{\lm}v_+^4\simeq\frac{1}{4\sqrt{3\lm}}m_-^2 \label{S1} \, ,
\end{align}
where we have neglected  $\mathcal{O}(\de,\vep)$. Note that $m_-^2\simeq27\lm v_+^4/16$. In closing,
we note that for consistency with the form of $V(\phi)$ shown in Fig.~\ref{fig:2}, in (\ref{S1}) we have inverted the extrema of integration as compared to the definition of $S_{1}$ given by CdL~\cite{Coleman:1980aw}. Note that our choice of the extrema of integration leads to a positive $S_{1}$.

\section{Bounce factor from  dimensional analysis and lifetime prerequisite}
\label{sec:5}

The bounce factor of a transition from dS to Minkowski in a 5-dim spacetime with  one compact dimension has been computed in~\cite{Antoniadis:2024ent} and is given by
\begin{equation}
B\simeq \frac{45\pi^3 R_{0} S_{1}^4}{\varepsilon^3} \Bigl[ 1-\hat{\kappa}_{5}+\mathcal{O}(\hat{\kappa}_{5}^2) \Bigr] \, ,
\end{equation}
whereas in a 5-dim spacetime with non-compact dimensions the result of CdL generalizes to
\begin{equation}
B \simeq \frac{2048\pi^2 S_{1}^5}{15\varepsilon^4} \Bigl[ 1-\frac{40}{21}\hat{\kappa}_{5}+\mathcal{O}(\hat{\kappa}_{5}^2) \Bigr] \,,
\end{equation}
where $\hat\kappa_5:=\kappa_5S_1^2/\vep$ is a dimensionless version of the five-dimensional reduced Planck length to the third power $\kappa_5$ introduced in~\cite{Antoniadis:2024ent}, controlling the gravitational correction to the bounce in the presence of gravity.
Next, we consider the gravity-decoupling limit, in which only the overall coefficients matter. In the thin-wall approximation,\footnote{We verify in Appendix~\ref{app:B} that the choice of the parameters discussed later in this section is consistent with the thin-wall approximation.} the scalar field $\phi$ varies on the wall, while the dimensionless scale factor $\rho$ of the spherical bubble (with radius $H^{-1}$) is considered nearly constant $\bar \rho$. Then, the  result for the compact dimension should be valid in the regime where the radius $R_{0}$ of the compact dimension is smaller than the radius $\bar{\rho} H^{-1} \simeq S_{1}/\varepsilon$ of the 3-sphere. When $R_{0}$ is larger than the latter, one should expect to reproduce the 5-dim non-compact result for the bounce, which is achieved by letting $R_{0} \simeq S_{1}/\varepsilon$. This is quantified in the negative mode analysis of the CdL instanton discussed in Sec.~\ref{sec:2}. Substituting 
\begin{equation}
\bar{\xi} = \frac{\bar{\rho}}{\sqrt{2}H} \qquad \text{and} \qquad \bar{\rho} = \frac{9\sqrt{2}HS_{1}}{3\varepsilon + 2\kappa_{5}S_{1}^2} \,\, ,
\end{equation}
into (\ref{ultima}) it is straightforward to see that in the limit $\kappa_{5}\rightarrow 0$ the compactification radius is constrained to satisfy
\begin{equation}
\bar{\xi} = \frac{3S_{1}}{\varepsilon} \qquad \Rightarrow \qquad R_{0} < \frac{\sqrt{3}S_{1}}{\varepsilon} \,\, .
\label{xi=S1vep}
\end{equation}
Putting all this together, the bounce configuration can be classified according to its two regimes: 
\begin{itemize}
\item $O(4)\times U(1)$ symmetric bounce in $\bfR^4\times S^1/\mathbf Z_2$, for which $R_0<\sqrt{3} \ S_1/\varepsilon$;
\item $O(5)$ symmetric bounce in $\bfR^5$, for which $R_0> \sqrt{3} \ S_1/\varepsilon$.
\end{itemize}
Before proceeding, we pause to clarify that even though for $R_0<\sqrt{3} \ S_1/\varepsilon$, the decay rate is described
by an effective 4D instanton, this decay rate describes transitions between 5D
vacua. Of critical interest for our investigation is that in the
intial vacuum the mass of the scalar
field is smaller than the
compactification radius, but after the transition the mass is bigger
than the compactification radius. Therefore, the transition should not be analyzed purely from a 4D perspective; namely, the field should be treated as 5D or else we must keep its
KK excitations.

In the following, we will use a unifying symbol $D$, which actually has already appeared in Eq.~\eqref{uno}: $D=4$ for the regime with the $O(4)\times U(1)$ symmetric bounce in $\bfR^4\times S^1/\mathbf{Z}_2$, and $D=5$ for the second regime with $O(5)$ symmetric bounce in $\bfR^5$. Note that $D$ is equal to the number of moduli of the bounce in each case. See Section~\ref{sec:2}.

The 5-dim action at the classical bounce on $\bfR^4\times S^1/\mathbf{Z}_2$ is found to satisfy
\begin{align}
\text{$O(4)\times U(1)$ symmetric regime}: &\quad R_0<\frac{\sqrt{3} \ S_1}{\varepsilon}, \quad 
B\simeq\frac{1400}{2}\frac{R_0S_1^4}{\varepsilon^3},
\label{stability-4dS1} \\
\text{$O(5)$ symmetric regime}: &\quad R_0>\frac{\sqrt{3} \ S_1}{\varepsilon}, \quad
B\simeq\frac{1350}{2}\frac{S_1^5}{\varepsilon^4},
\label{stability-5d}
\end{align}
where the factor of two in the denominator in both cases comes from the orbifold $S^1/\mathbf{Z}_2$ describing a line interval with two ends.

In anticipation to our calculations we recall that the AdS $\to$ dS transition at $z_c$ demands the lifetime of the meta-stable vacuum to be about half the age of the universe, i.e.,
\begin{align}
\label{tau-value}
\tau \sim 3.3 \times 10^{32}~{\rm eV}^{-1} \, .
\end{align}
In addition, we require the mass ratio $m_-/m_+$ and the vacuum energy difference $\vep$ of our sixth order potential $V$ to satisfy~\cite{Anchordoqui:2023woo}
\begin{align}
\frac{m_-}{m_+} \gtrsim 5 \label{m+/m-5} 
\end{align}
and
\begin{align}
\vep\ll\Lambda_5\sim(20\,\mbox{meV})^5 \label{vepvsLm} \,.
\end{align}
Moreover, $m_+$ should be much smaller than the compactification scale $R_0^{-1}$ and bigger than the 5D cosmological constant $\Lambda_5^{1/5}$, which fixes its value around $m_+\sim100$\,meV.
Now, (\ref{m+/m-5}) yields $\de\lesssim0.01$ (see Eq.~\eqref{parameters-in-delta2}).
The lifetime of the false vacuum  is related to the decay rate (\ref{uno}) by
\begin{align}
\tau\sim\left(\frac{\Ga}{V_{D-1}}\right)^{-1/D} \,.  \label{tauGaV}
\end{align}
The relation \eqref{tauGaV} can be rewritten as 
\begin{align}
B\sim D\ln \left(\tau A^{1/D} \right) \,.  \label{BDlnA} 
\end{align}
To proceed further, we need a more explicit expression of the one-loop factor $A$, which is defined by~\cite{Callan:1977pt}
\begin{align}\label{detratio}
A=\left(\frac{B}{2\pi}\right)^{D/2}\left|\frac{\det'O_{\text{B}}}{\det O_{\text{FV}}}\right|^{-1/2},
\end{align}
where the factor $(B/(2\pi))^{D/2}$ comes from the integration measure on the moduli space. $O_{\text{B}}$ and $O_{\text{FV}}$ are the differential operators that define the eigenmode equations for linear fluctuations around the bounce $\phi_\rB$ and the false vacuum $v_+$, respectively, (see also Sec.~\ref{sec:2})
\begin{align}
O_{\text{B}}&=-\triangle_d+V''(\phi_\rB)\,, \label{OB} \\
O_{\text{FV}}&=-\triangle_d+V''(v_+) \,. \label{OFV}
\end{align}
The primed determinant $\det'O_{\text{B}}$ is the determinant of $O_{\text{B}}$ with $D$ zero modes of spin 1 removed, and $\det O_{\text{FV}}$ is the full determinant of $O_{\text{FV}}$. Recall that each zero mode of $O_\rB$ corresponds to each modulus of the bounce and hence the number of the zero modes is $D$. Since $O_\rB$ and $O_{\mathrm{FV}}$ have mass dimension 2, the ratio ${\det'O_{\text{B}}}/{\det O_{\text{FV}}}$ has dimension $-2D$ and hence $A$ has dimension $D$. See Appendix~\ref{sec:app1} for more details.

In the following, instead of using an analytic expression of $A$, we exploit the scaling factor of $A$ that carries the mass dimension $D$. We can show (see Appendix~\ref{sec:app2}) that the scale of the determinant ratio is carried by $(\sqrt{\lm}v_+^2)^D$, which is proportional to $m_-^D$ by \eqref{V0''(0)}. Therefore, we can express $A$ as 
\begin{align}\label{Aina}
A\sim\left(\frac{B}{2\pi}\right)^{D/2} \ a \ m_-^D\,, 
\end{align}
where $a$ is the dimensionless factor (note that $B$ is dimensionless). We may then rewrite Eq.~\eqref{BDlnA} into a transcendental equation of $B$,
\begin{align}
B\sim\frac{D}{2}\ln\frac{B}{2\pi}+D\ln(\tau \ m_- \ a)\,.   \label{Bequation}
\end{align}
A remark is in order. Though $m_-$ was chosen to carry the dimension of $A$, it is possible that other parameters of dimension 1 such as $m_+$ carry part or all of the mass dimension, and also that some power of $\delta$ may enter $A$ as an overall factor. 
Such ambiguity amounts to the ambiguity about $a$.

Though concrete expression and numerical value of $a$ are difficult to obtain, we can actually demonstrate that this does not contribute a big factor, thus not invalidating our discussion in the rest, as we will see later on. 
For the meantime, we will just set $a=1$ for simplicity. Using our fiducial value
$m_- \sim 500~\text{meV}$,
we can solve the equation \eqref{Bequation} numerically to obtain
\begin{align}
\text{$O(4)\times U(1)$ symmetric regime}: &\quad B \sim 304\,, \label{BD=4}
\end{align}
and
\begin{align}
\text{$O(5)$ symmetric regime}: &\quad B \sim 380\,. \label{BD=5}
\end{align}

Next, we investigate the consequences of \eqref{BD=4} and \eqref{BD=5} by combining them  with the expressions of $B$ given in \eqref{stability-4dS1} and \eqref{stability-5d} to obtain
\begin{align}
  \text{$O(4)\times U(1)$ symmetric regime}: &\quad \frac{S_1}{\varepsilon}\sim\left(\frac{300}{700}\frac{1}{R_0\varepsilon}\right)^{1/4}, \label{S1vep-4d} 
\end{align}
and
\begin{align}
\text{$O(5)$ symmetric regime}: &\quad \frac{S_1}{\varepsilon}\sim\left(\frac{380}{675}\frac{1}{\varepsilon}\right)^{1/5}. \label{S1vep-5d}
\end{align}
For concreteness, we assume $
R_0 \sim 10\,\mu\text{m}\sim8\,\text{eV}^{-1}$  and $\varepsilon\lesssim(2\,\mbox{meV})^5$,
yielding
\begin{align}
R_0\varepsilon \lesssim 10^{-12}\,\text{eV}^4.    
\label{R0less}
\end{align}
Combining (\ref{R0less}) with the expression of $S_1/\varepsilon$ in \eqref{S1vep-4d} and \eqref{S1vep-5d}, we estimate a bound on $S_1/\varepsilon$ for both regimes:
\begin{itemize}[noitemsep,topsep=0pt]
\item For the $O(5)$ symmetric regime, we obtain
\begin{align}
\frac{S_1}{\varepsilon}\sim\left(\frac{380}{675}\frac{1}{\varepsilon}\right)^{1/5}\gtrsim445\,\text{eV}^{-1},
\end{align}
which is incompatible with the definition of the 5-dim non-compact configuration of the bound,
\begin{align}
\frac{S_1}{\varepsilon}<\frac{R_0}{\sqrt{3}}\sim 6~\text{eV}^{-1}.
\end{align}
\item For the $O(4)\times U(1)$ symmetric regime, we obtain
\begin{align}\label{S1epO4}
\frac{S_1}{\varepsilon}\sim\left(\frac{300}{700}\frac{1}{R_0\varepsilon}\right)^{1/4}\gtrsim720\,\text{eV}^{-1},
\end{align}
which turns out to automatically satisfy the stability condition, 
\begin{align}
\frac{S_1}{\varepsilon}>\frac{R_0}{\sqrt{3}}\sim 6~\text{eV}^{-1}.
\end{align}
Here we come back to the effect of the parameter $a$ in Eq.~\eqref{Bequation}. As $a$ increases, the solution $B$ increases. Concretely, as $a$ varies from $10^{-4}$ to $10^4$, the solution $B$ varies from 260 to 340 for the $O(4)\times U(1)$ symmetric bounce ($D=4$). In accord, the value of $S_1/\vep$ varies from 700 to 740, still satisfying the stability condition. $S_1/\vep$ saturates the stability condition when $a$ goes down to order $\mathcal{O}(10^{-31})$.
\end{itemize}
All in all, a 5-dim instanton with a compact dimension can naturally explain an AdS $\to$ dS transition at $z \sim 2$. 

Armed with our finding we now provide an assessment on the size of
  the bubble and the possiblity of having bubble collisions. The size
  of the bubble is estimated in Eq.~\eqref{stability-4dS1} and is given by $\bar \xi = 3 S_1/\varepsilon$. We demand that
$\bar \xi$ is bigger than a minimum length which is proportional to the
compactification scale to avoid the negative mode
instability. However, we stress that this is a lower bound for the
size of the bubble. To find an upper bound, we first note that once the
lifetime of the 5D false vacuum $\tau$ and the true-vacuum mass $m_-$
are fixed (see Eqs.~\eqref{tau-value} and \eqref{Bequation}), the value of $B$ is essentially determined (see Eq.~\eqref{BD=4}).
Now, we can conclude that $S_1/\varepsilon$
as given by Eq.~\eqref{S1vep-4d}  corresponds to the probability of a decay at
$z=2$. As previously noted, Eq.~\eqref{S1epO4}  provides a lower bound for $S_1/\varepsilon$, which is $720~{\rm
  eV}^{-1}$. But of course we can increase the size of the bubbles by
decreasing $\varepsilon$ while
keeping fixed the probability determined by $B$. From  Eq.~\eqref{S1vep-4d} this means that $S_1/\varepsilon
\propto \varepsilon^{-1/4}$. This in turn means that parametrically we can make the size
of the bubble as large as the size of the universe. Obviously, the
limit must be taken with $S_1$ and $\varepsilon$ going to zero with
different rates and this is dictated by the condition that $B$ remains constant.

In summary, the size of the bubble is a parameter that can be
assessed in our model. The size is not fixed, but for fixed probability
it depends parametrically on $\varepsilon$. The size then can be
arbitrarily large parametrically, and as a matter of fact it can be as large as the
size of the universe at the time of the transition.

In practice, if the bubble is very large we expect no bubble collisions. As we
start decreasing the size, the dynamics becomes more and more important 
because collisions will start playing some role. Actually, there may
be a critical size for which the process destabilizes.  Indeed, if the bubble size $\bar\xi$ is much smaller than the horizon scale $\sim 1/H_0$, one may need a dedicated analysis of dynamics of the bubbles, which is beyond the scope of this work.

\paragraph{Decoupling limit of gravity}
Let us justify our assumption of neglecting the gravitational corrections, namely $\hat\kappa_5\sim0$. 
We consider the regime with the $O(4)\times U(1)$ symmetric bounce. The first relation in \eqref{S1epO4} can be rewritten as
\begin{align} \label{S12/vep}
\frac{S_1^2}{\vep}\sim\left(\frac{3\vep}{7R_0}\right)^{1/2} \,.
\end{align}
The gravitational correction becomes important when $\hat\kappa_5 \gtrsim 1$, which is equivalent to 
\begin{align}
M_*^3 \lesssim \frac{S_1^2}{\vep} \sim\left(\frac{3\vep}{7R_0}\right)^{1/2} \,,
\end{align}
where $M_*$ is the species scale (see Sec.~\ref{sec:1}), related to $\kappa_5$ as $\kappa_5=M_*^{-3}$, and we used \eqref{S12/vep} in the second step. This can be further rewritten in terms of the ratio of the vacuum energy scale $\vep^{1/5}$ to the Planck mass $M_*$,
\begin{align} \label{decoupling-lowerbound}
\frac{\vep^{1/5}}{M_*} \gtrsim \left(\frac{7}{3}R_0M_*\right)^{1/5} \sim 10^4 \,,
\end{align}
where we used $R_0\sim 8\,\text{eV}$ and $M_*\sim 10^9\,\text{GeV}$. However, this big lower bound means the breakdown of the effective theoretical treatment of our 5-dim action of $\phi$, and it is also impossible in the context of our analysis due to the upper bound on $\vep$ given by \eqref{vepvsLm}.

\section{Conclusions}
\label{sec:6}

We have particularised the analysis of
vacuum decay in the presence of a  compact dimension presented elsewhere~\cite{Antoniadis:2024ent} to validate the hypotheses of an AdS $\to$ dS transition driven by the Casimir forces of fields inhabiting the incredible bulk of the dark dimension scenario. Such a transition was proposed in~\cite{Anchordoqui:2023woo} to explain a late time ($z \sim 2$) rapid sign-switching cosmological constant, which can significantly improved the fit to observational data and resolves the $H_0$ and $S_8$ tensions~\cite{Akarsu:2023mfb}.

We adopted the Callan-Coleman-de Luccia formalism for calculating the
transition probability within the thin-wall approximation. We have
shown that the Euclidean bounce configuration that drives the vacuum
decay cannot be realised by a quartic potential and we have used a
minimal sixth order one. We have also shown that distinctive features
of the required vacuum decay to accommodate the AdS $\to$ dS
transition are inconsistent with a 5-dim non-compact description of
the instanton, for which the bounce is $O(5)$ symmetric, and instead
call for 5-dim instanton with a compact dimension, for which the
bounce is $O(4)\times U(1)$ symmetric. It should be
  emphasized that in the intial vacuum the mass of the scalar
field undergoing the transition is smaller than the
compactification radius, but after the transition the mass is bigger
than the compactification radius. This implies that the scalar field is 5D, although the probability that
characterises the transition is described by an effective 4D instanton.

We end by noting that the Dark Energy Spectroscopic Instrument (DESI) Collaboration recently measured a tight relation between $H_0$ and the distance to the Coma cluster~\cite{Said:2024pwm}. More recently, it was noted that the inverse distance ladder of the Hubble diagram from the DESI relation combined with $H_0$ as
 determined by CMB observations with $\Lambda$CDM extrapolation leads to an Earth-Coma distance $d_{\rm EC} = (111.8 \pm 1.8)~{\rm Mpc}$, which is $4.6\sigma$ larger than the value $d_{\rm EC} = (98.5 \pm 2.2)~{\rm Mpc}$ obtained from calibrating the absolute magnitude of SNe Ia with the Hubble Space Telescope distance ladder~\cite{Scolnic:2024hbh}. Needless to say, the canonical value  $95 \lesssim d_{\rm EC}/{\rm Mpc} \lesssim 100$ is consistent with the $H_0$ measurement by SH0ES~~\cite{Riess:2021jrx,Murakami:2023xuy}. It is hard to imagine how Coma could be located as far as  $>110~{\rm Mpc}$. By extending the Hubble diagram to Coma, DESI data point to a momentous conflict between our knowledge of local distances and cosmological expectations from $\Lambda$CDM extrapolations. A late time rapid sign-switching cosmological constant based on the ideas discussed in this paper would provide a resolution of the Earth-Coma-distance conflict.

\section*{Acknowledgements}

The work of L.A.A. is supported by the U.S. National Science
Foundation (NSF Grant PHY-2412679). I.A. is supported by the Second
Century Fund (C2F), Chulalongkorn University. D.B., A.C. and H.I. have been supported by Thailand NSRF via PMU-B, grant number B37G660014 and B13F670063.

\appendix

\section{One-loop determinant ratio}
\label{sec:app}
In this Appendix, we give some properties of the determinants of differential operators we used in the main part and describe how the scale of the one-loop determinant ratio is determined.

\subsection{Determinants}
\label{sec:app1}
We first summarise definitions and properties of the determinants. We recall the definitions:
\begin{align}
O_{\text{B}}&=-\triangle_d+V''(\phi_\rB)\,, \label{OB-app} \\
O_{\text{FV}}&=-\triangle_d+V''(v_+) \,. \label{OFV-app}
\end{align}
We follow the notations given in Sec.~\ref{sec:2}. 

\subsubsection{$O(d)$ symmetric bounce}
In this case, the determinant $\det O_\rB$ is given by
\begin{align}\label{det-O(d)-app}
\det O_\rB=\prod_{\ell \geq 0}(\det \tilde O_{\rB,\ell})^{\mathfrak{d}_{\ell,d}} \,, \qquad
\det\tilde O_{\rB,\ell}=\prod_{h_\ell \in \sig(\tilde O_{\rB,\ell})}h_\ell \,,
\end{align}
where the power $\mathfrak{d}_{\ell,d}$ reflects the multiplicity of each eigenvalue of $\tilde O_{\rB,\ell}$. 

Since $\tilde O_{\rB,1}$ has one zero eigenvalue, the spin-1 part $\det \tilde O_{\rB,1}$ becomes zero. We therefore modify this into $\det{}' \tilde O_{\rB,1}$ by removing the zero mode,
\begin{align}
\det{}'\tilde O_{\rB,\ell}:=\prod_{\substack{h_\ell \in \sig(\tilde O_{\rB,\ell}) \\ h_\ell \neq 0}}h_\ell \,.
\end{align}
The total determinant after this replacement is denoted with prime by
\begin{align}\label{det-O(d)app}
\det{}'O_\rB:=\det\tilde O_{\rB,0} \cdot (\det{}'\tilde O_{\rB,1})^d\prod_{\ell=2}^\infty(\det \tilde O_{\rB,\ell})^{\mathfrak{d}_{\ell,d}} \,.
\end{align}

We also need $\det O_{\mathrm{FV}}$ since the decay rate is expressed with the determinant ratio as \eqref{detratio}~\cite{Callan:1977pt}. The corresponding radial differential operator $\tilde O_{\mathrm{FV},\ell}$ in the spin $\ell$ representation of $O(d)$ is obtained by replacing $V''(\phi_\rB)$ in $\tilde O_{\rB,\ell}$ \eqref{radeigenmodeeq-d2} by $V''(v_+)=m_+^2$. The determinant $\det O_{\mathrm{FV}}$ is defined in the same way,
\begin{align}\label{detFV-O(d)app}
\det O_{\mathrm{FV}}=\prod_{\ell \geq 0}(\det \tilde O_{\mathrm{FV},\ell})^{\mathfrak{d}_{\ell,d}} \,, \qquad
\det\tilde O_{\mathrm{FV},\ell}=\prod_{h_\ell \in \sig(\tilde O_{\mathrm{FV},\ell})}h_\ell \,,
\end{align}
where $\sig(\tilde O_{\mathrm{FV},\ell})$ is the set of all eigenvalues of $\tilde O_{\mathrm{FV},\ell}$. The determinant ratio then reads
\begin{align}
\frac{\det{}'O_\rB}{\det O_{\mathrm{FV}}}=\frac{\det\tilde O_{\rB,0}}{\det\tilde O_{\mathrm{FV},0}}\left[\frac{\det{}'\tilde O_{\rB,1}}{\det\tilde O_{\mathrm{FV},1}}\right]^d\prod_{\ell=2}^\infty\left[\frac{\det\tilde O_{\rB,\ell}}{\det\tilde O_{\mathrm{FV},\ell}}\right]^{\mathfrak{d}_{\ell,d}} \,,
\end{align}
which makes sense because $\tilde O_{\mathrm{FV},\ell}$ is positive definite for any $\ell$. Since $\det{}'\tilde O_{\rB,1}$ is missing one eigenvalue and the other determinant ratios $(\ell\neq 1)$ are dimensionless, the total ratio $\det{}'O_\rB/\det O_{\mathrm{FV}}$ has mass dimension $-2d$. 

A remark is that $\tilde O_{\mathrm{FV},0}$ might seem to have eigenvalue $m_+^2$ with a non-vanishing, constant eigenfunction, but it is not true because this eigenfunction contradicts the boundary condition in \eqref{radeigenmodeeq-d1}.

\subsubsection{$O(d-1)\times U(1)$ symmetric bounce}
In this case, the determinant is given by
\begin{align}\label{det-O(d-1)app}
&\det O_\rB=\prod_{\ell=0}^\infty~(\widehat{\det}\,\tilde O_{\rB,\ell})^{\mathfrak{d}_{\ell,d-1}} \,, \\
&\widehat{\det}\,\tilde O_{\rB,\ell}:=
\prod_{h_\ell \in \sig_{\mathrm{KK}}(\tilde O_{\rB,\ell})}~\prod_{s=0}^\infty\left(h_\ell+\frac{s^2}{R_0^2}\right) \,.
\end{align}
Since $\tilde O_{\rB,1}$ has one zero eigenvalue, the spin-1 part $\widehat{\det}\,\tilde O_{\rB,1}$ is zero. We therefore modify this into  $\widehat{\det}{}'\,\tilde O_{\rB,1}$ by removing the zero mode,
\begin{align}
\widehat{\det}{}'\,\tilde O_{\rB,\ell}=\prod_{\substack{h_\ell \in \sig_{\mathrm{KK}}(\tilde O_{\rB,\ell}) \\ h_\ell \neq 0}}~\prod_{s=0}^\infty\left(h_\ell+\frac{s^2}{R_0^2}\right) \,.
\end{align}
The total determinant under this replacement is given with prime by
\begin{align}\label{detFV-O(d-1)app}
&\det{}' O_\rB=\widehat{\det}\,\tilde O_{\rB,0} \cdot (\widehat{\det}{}'\,\tilde O_{\rB,1})^{d-1} \prod_{\ell=2}^\infty~\widehat{\det}\,\tilde O_{\rB,\ell} \,.
\end{align}

Let us next consider $\det O_{\mathrm{FV}}$. The corresponding radial differential operator $\tilde O_{\mathrm{FV},\ell}$ in the spin $\ell$ representation of $O(d-1)$ is obtained by replacing $V''(\phi_\rB)$ in $\tilde O_{\rB,\ell}$ \eqref{eigenpsi-O(d-1)2} by $V''(v_+)=m_+^2$. The determinant $\det O_{\mathrm{FV}}$ is defined in the same way,
\begin{align}\label{det-O(d)}
&\det O_{\mathrm{FV}}=\prod_{\ell=0}^\infty~(\widehat{\det}\,\tilde O_{\mathrm{FV},\ell})^{\mathfrak{d}_{\ell,d-1}} \,, \\
&\widehat{\det}\,\tilde O_{\mathrm{FV},\ell}:=
\prod_{h_\ell \in \sig(\tilde O_{\mathrm{FV},\ell})}~\prod_{s=0}^\infty\left(h_\ell+\frac{s^2}{R_0^2}\right) \,,
\end{align}
where $\sig(\tilde O_{\mathrm{FV},\ell})$ is the set of all eigenvalues of $\tilde O_{\mathrm{FV},\ell}$. 
The determinant ratio then reads
\begin{align}
\frac{\det{}'O_\rB}{\det O_{\mathrm{FV}}}=\frac{\widehat{\det}\,\tilde O_{\rB,0}}{\widehat{\det}\,\tilde O_{\mathrm{FV},0}}\left[\frac{\widehat{\det}{}'\,\tilde O_{\rB,1}}{\widehat{\det}\,\tilde O_{\mathrm{FV},1}}\right]^{d-1}\prod_{\ell=2}^\infty\left[\frac{\widehat{\det}\,\tilde O_{\rB,\ell}}{\widehat{\det}\,\tilde O_{\mathrm{FV},\ell}}\right]^{\mathfrak{d}_{\ell,d-1}} \,,
\end{align}
which makes sense because $\tilde O_{\mathrm{FV},\ell}$ is positive definite for any $\ell$. Since $\widehat{\det}{}'\,\tilde O_{\rB,1}$ is missing one eigenvalue and the other determinant ratios $(\ell\neq 1)$ are dimensionless, the total ratio $\det{}'\,O_\rB/\det O_{\mathrm{FV}}$ has mass dimension $-2(d-1)$. 

As in the $O(d)$ symmetric case, $m_+^2$ is not an eigenvalue of $\tilde O_{\mathrm{FV},0}$ because its nonvanishing, constant eigenfunction contradicts the boundary condition in \eqref{eigenpsi-O(d-1)1}.

\subsection{Scaling behaviour}
\label{sec:app2}
We start from the EoM for the bounce, which reads
\begin{align}
\ddot\phi_\rB(\xi)+\frac{D-1}{\xi}\dot\phi_\rB(\xi)-V'(\phi_\rB)=0 \,,
\end{align}
where $V(\phi)$ is the sixth order potential \eqref{phi6_V}, and $D=4$ for the $O(4)\times U(1)$ symmetric bounce in $\bfR^4\times S^1/\mathbf{Z}_2$ and $D=5$ for the $O(5)$ symmetric bounce in $\bfR^5$.
Let us rescale the radial coordinate $\xi$, the parameter $\vep$, and the bounce $\phi_\rB$ to make them dimensionless,
\begin{align}\label{rescaling}
z=\sqrt{\lm}v_+^2\xi\,, \quad \vphi_\rB(z):=\frac{\phi_\rB(\xi)}{v_+} \,, \quad \hat\vep:=\frac{\vep}{\lm v_+^6}\,.
\end{align}
In terms of the dimensionless quantities, the EoM becomes
\begin{align}
\frac{d^2\vphi_\rB}{dz^2}+\frac{3}{z}\frac{d\vphi_\rB}{dz}=\hat V'_0(\vphi_\rB)+\frac{\hat\vep}{1-r} \,,
\end{align}
where $\hat V'_0(\vphi_\rB)=\vphi_\rB(\vphi_\rB-1)(\vphi_\rB-r)^3$.
Since this EoM contains only dimensionless parameters $r$ and $\hat\vep$, the parameters in the dimensionless bounce $\vphi_\rB$ are only $r$ and $\hat\vep$.

Let us rescale the eigenmode equation $O_\rB\eta=h\eta$ around the $O(4)\times U(1)$ symmetric bounce with \eqref{rescaling} together with $\hat y=\sqrt{\lm}v_+^2y$. The result is
\begin{align}
\left[ -\frac{\partial^2}{\partial z^2}-\frac{3}{z}\frac{\partial}{\partial z}-\frac{\triangle_{S^{d-2}}}{z^2}+\hat V''(\vphi_\rB)+\frac{\pd^2}{\pd\hat y^2} \right] \eta = \frac{h}{\lm v_+^4} \eta \,.
\end{align}
Since the parameters on its left hand side are only $r,\hat\vep$, the eigenvalue $h/(\lm v_+^4)$ is a dimensionless quantity with these parameters. Therefore, the dimension of eigenvalue $h$ is carried by $\lm v_+^4$, which is proportional to $m_-^2$ by \eqref{parameters-in-delta}. Concretely, each eigenvalue can be expressed as (see also Sec.~\ref{subsec:O(4)U(1)})
\begin{align}
m_-^2 \left( \frac{h_\ell}{m_-^2} + \frac{s^2}{m_-^2R_0^2} \right) \,,
\end{align}
where the term with the round bracket is dimensionless and $h_\ell/m_-^2$ is a function of $r$ and $\hat\vep$.
This argument goes in a parallel manner in the case of the $O(5)$ symmetric bounce. It is also obvious that the same argument holds for $O_{\mathrm{FV}}\eta=h\eta$ around the false vacuum. Therefore, in both cases, any eigenvalue of the eigenmode equations can be written as a product of the factor $m_-^2$ and a dimensionless quantity depending on $r,\hat\vep$ and $m_-R_0$.

Therefore, the mass dimension $-2D$ of the determinant ratio $\det{}'\,O_\rB/\det O_{\mathrm{FV}}$, where $D$ is the number of the moduli of the bounce solution, is carried by the factor $m_-^{-2D}$, and hence the one-loop factor $A$ in \eqref{detratio} can be expressed as \eqref{Aina},
\begin{align}\label{Aina-app}
A\sim\left(\frac{B}{2\pi}\right)^{D/2} \ a \ m_-^D\,,
\end{align}
where $a$ is a function of dimensionless quantities $r,\hat\vep$ and $m_-R_0$. Note that $A$ should exist in the limit $r\to-1/2$ $(\de\to0)$ and $\hat\vep\to0$.

A remark is in order. The rescaling \eqref{rescaling} is not the unique one, but we may further multiply powers of dimensionless factors. For example, we can adopt the rescaling so that the dimension 2 of the eigenvalues is carried by $m_+^2$ instead of $m_-^2$. For the one-loop factor $A$, this change of the scale factor can be absorbed into the change in the dependence of the dimensionless parameter $a$ on the mass ratio $m_+^2/m_-^2\sim\de$. 

\section{Comment on the thin-wall approximation}
\label{app:B}

In this Appendix, we will make a brief review of the thin-wall approximation following~\cite{Coleman:1977py}, and verify that the approximation is consistent with our choice of the parameters discussed in Sec.~\ref{sec:5}. As in the other parts, we use the symbol $D$ as follows: $D=d$ in the case with the $O(d)$ symmetric bounce and $D=d-1$ in the case with the $O(d-1)\times U(1)$ symmetric bounce. In either case, the bounce configuration is a spherically symmetric configuration $\phi_\rB(\xi)$ characterised by
\begin{align}
\lim_{\xi\to\infty}\phi_\rB(\xi)=\phi_+, \quad
\dot\phi_\rB(\xi=0)=0,
\end{align}
where $\xi$ is the radial coordinate of the Euclidean spacetime $\bfR^D$ defined by
\begin{align}
\xi=\sqrt{\tau^2+(x^1)^2+\cdots+(x^{D-1})^2},
\end{align}
and $\dot\phi_\rB=d\phi_\rB/d\xi$.\footnote{Note that the bounce configuration is independent of the compact direction in the $O(d-1)\times U(1)$ symmetric case.}
The bounce $\phi_\rB$ satisfies the equation of motion
\begin{align}
\frac{d^2\phi_\rB}{d\xi^2}+\frac{D-1}{\xi}\frac{d\phi_\rB}{d\xi}-V'(\phi_\rB)=0.
\end{align}
We can therefore consider $\phi_B$ describing a motion with time $\xi\geq 0$ in the flipped potential $-V(\phi)$ under the initial condition $\dot\phi_\rB(0)=0$. Note that $v_+<\phi_\rB(0)<v_-$, where $v_-$ $(v_+)$ is the true (false) vacuum.  

Let us consider the initial position $\phi_\rB(0)$. Ref.~\cite{Coleman:1977py} considered a particular case where the initial position is sufficiently close to the true vacuum, $\phi_\rB(0) \sim v_-$. Let us first look at the behaviour of $\phi_\rB$ near the true vacuum where the potential can be approximated as
\begin{align}
V(\phi)\simeq\frac{m_-^2}{2}(\phi-v_-)^2.
\end{align}
The equation of motion then becomes
\begin{align}
\frac{d^2\phi_\rB}{d\xi^2}+\frac{D-1}{\xi}\frac{d\phi_\rB}{d\xi}-m_-^2(\phi_\rB-v_-)=0,  
\end{align}
which yields an exact solution,
\begin{align}
\phi_\rB(\xi)-v_- = [\phi_\rB(0)-v_-]f_D(m_-\xi),
\end{align}
where $f_D$ is a monotonically increasing function defined by\footnote{The function $I_\nu(z)$ is the modified Bessel function of the first kind with order $\nu$.}
\begin{align}
f_D(z):=\Ga\left(\frac{D}{2}\right)\left(\frac{z}{2}\right)^{1-\frac{D}{2}}I_{\frac{D}{2}-1}(z),
\end{align}
which satisfies $f_D(z)\to 1$ as $z\to 0$ and increases exponentially for large $z$. 
As $\phi_\rB(0)-v_-$ gets smaller, the growth rate of $f_D(m_-\xi)$ will get more suppressed
so that it will take longer time in the vicinity of $v_-$ until it increases exponentially around some time $\bar\xi$. Therefore, by taking $\phi_\rB(0)$ to be very close to $v_-$, $\phi_\rB$ stays for very long time until some sufficiently large time $\bar\xi$ such that the friction $(D-1)\dot\phi_\rB/\xi$ can be negligible after $\bar\xi$. Since $\phi_\rB$ moves from $\phi_\rB(0)\sim v_-$ to $v_+$ without friction after $\bar\xi$, energy conservation implies that $\phi_\rB(0)$ should be taken such that $V(\phi_\rB(0))\sim V(v_+)$ to avoid going beyond the false vacuum (overshoot) and not reaching it (undershoot). In summary, we require
\begin{align}\label{phirB(0)-conds}
\phi_\rB(0)\sim v_-, \quad V(\phi_\rB(0))\sim V(v_+).
\end{align}
These are compatible if the energy difference of the two vacua are very small,
\begin{align}\label{V+simV-}
V(v_+) \sim V(v_-).
\end{align}

Actually, we can reverse the argument in the last paragraph: one can show~\cite{Coleman:1977py} that whether a potential $V$ satisfies the condition \eqref{V+simV-} or not, there exists an initial position $\phi_\rB(0)$ such that $\phi_\rB$ can reach the false vacuum $v_+$, by paraphrasing the arguments above about $\phi_\rB$ near the true vacuum and about the overshoot/undershoot combined with continuity of the motion. In particular with a potential with a small vacuum energy difference \eqref{V+simV-}, $\phi_\rB(0)$ thus obtained should be in the vicinity of the true vacuum $v_-$.

\begin{figure}[htb!]
\centering
\includegraphics[scale=0.45]{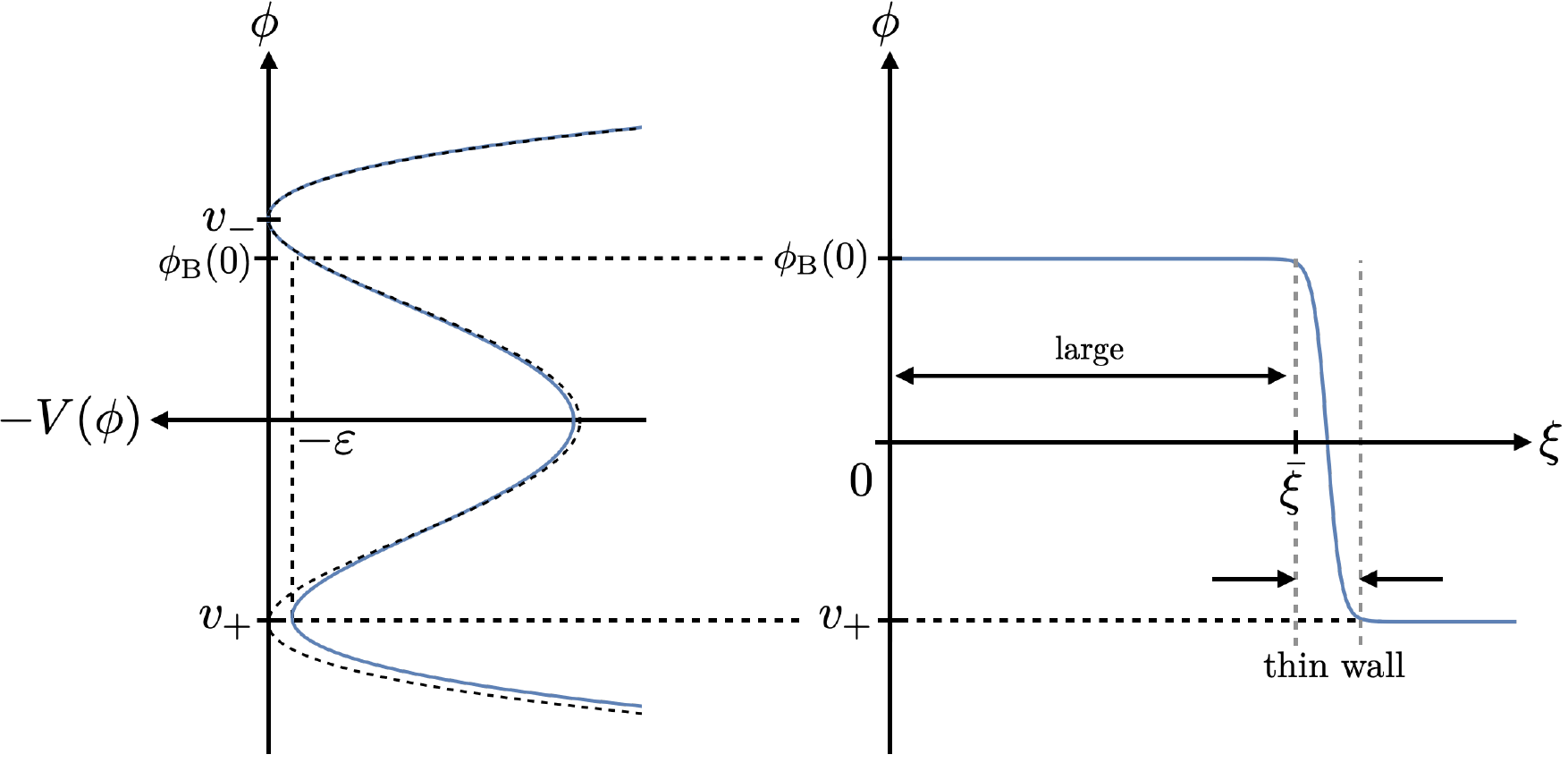} 
\caption{A flipped potential $-V(\phi)$ (left) and $\phi_\rB$ in the thin-wall approximation (right). The initial position $\phi_\rB(0)$ is chosen such that $V(\phi_\rB(0))\sim V(v_-)$. In the thin-wall approximation, the two vacua have a very small energy difference, so that $\phi_\rB(0)$ is located very near the true vacuum $v_-$. The closer $\phi_\rB(0)$ is to $v_-$, the larger $\bar\xi$ becomes. The friction is negligible after large $\bar\xi$.}
\label{fig:thinwall}
\end{figure}

Now, the scenario of the thin-wall approximation with a potential satisfying \eqref{V+simV-} is the following: $\phi_\rB$ starts at $\phi_\rB(0)$ near the true vacuum $v_-$ under the conditions \eqref{phirB(0)-conds}, stays near $v_-$ until a large time $\bar\xi$, rapidly rolls down and climb up the flipped potential $-V$ with a much smaller time scale than $\bar\xi$, and then reaches the false vacuum $v_+$ taking long time again. Fig.~\ref{fig:thinwall} shows a flipped potential $-V(\phi)$ satisfying \eqref{V+simV-} (left) and $\phi_\rB$ in the thin-wall approximation (right). As seen there, ``thin-wall'' implies that the time scale of the motion from $\phi_\rB(0)\sim v_-$ to $v_+$ is much smaller than $\bar\xi$.

Let us focus on the 6th order potential \eqref{phi6_V} discussed in Sec.~\ref{sec:4} and \ref{sec:5} to find a quantitative expression of the thin-wall approximation. 
As indicated above, it is sufficient to impose that $\bar\xi$ should be much larger than the largest time scale during the change of $\phi_\rB$ from $v_-$ to $v_+$.
$\phi_\rB$ departs from the true vacuum with exponentially large velocity, rolling down and climbs up the flipped potential $-V$ quickly without friction. After climibing up $-V$, the motion of $\phi_\rB$ near the false vacuum is approximately $\sim e^{-m_+\xi}$ since the friction is negligible, which has the time scale $\sim m_+^{-1}$ and can contribute the largest time scale since in the potential \eqref{phi6_V}, the mass $m_+$ around the false vacuum of the potential is required to be substantially smaller than that around the true vacuum $m_-$ (see \eqref{m+/m-5}). Therefore, as the thin-wall condition, it is sufficient to impose
\begin{align}
\bar\xi \gg m_+^{-1},
\end{align}
which can be rewritten with the expression of $\bar\xi$ given in \eqref{xi=S1vep} as
\begin{align}\label{S1vep103}
\frac{S_1}{\vep} \gg \frac{m_+^{-1}}{3}.
\end{align}
On the other hand, in the case of the $O(4)\times U(1)$ symmetric bounce for which $D=4$, we have another lower bound \eqref{S1epO4}:
\begin{align}\label{S1epO4-app}
\frac{S_1}{\vep}\gtrsim720\,\text{eV}^{-1}.
\end{align}
One can then observe that $S_1/\vep$ with this lower bound can automatically satisfy the thin-wall condition \eqref{S1vep103} as long as 
\begin{align}
m_+^{-1} \ll 2160\,\text{eV}^{-1} \quad \Longrightarrow \quad
m_+ \gg 0.5\,\text{meV}.
\end{align}
This is satisfied by our choice of the parameters in Sec.~\ref{sec:5}.

\providecommand{\href}[2]{#2}\begingroup\raggedright\endgroup

%\bibliography{Vacuum_Decay_2}

\end{document}